\documentstyle[12pt]{article}
\catcode`\@=11
\@addtoreset{equation}{section}
\def\section{\@startsection{section}{1}{\z@}{3.5ex plus 1ex minus .2ex}{2.3ex
plus .2ex}{\bf}}
\def\subsection{\@startsection{subsection}{2}{\z@}{3.25ex plus 1ex minus
.2ex}{1.5ex plus .2ex}{\it}}
\def\subsubsection{\@startsection{subsubsection}{2}{\z@}{3.25ex plus 1ex minus
.2ex}{1.5ex plus .2ex}{\it}}

\def\anp#1#2#3{{\it Ann. of Phys. (N.Y.)} {\bf #1} (19#3) #2 }
\def\phl#1#2#3{{\it Phys. Lett.} {\bf #1} (19#3) #2}
\def\jphys#1#2#3{{\it J. Phys.} {\bf #1} (19#3) #2}

\def\prl#1#2#3{{\it Phys. Rev. Lett.} {\bf #1} (19#3) #2}
\def\rmp#1#2#3{{\it Rev. Mod. Phys.} {\bf #1} (19#3) #2}
\def\phr#1#2#3{{\it Phys. Rep.} {\bf #1} (19#3) #2}
\def\pr#1#2#3{{\it Phys. Rev.} {\bf #1} (19#3) #2}
\def\nup#1#2#3{{\it Nucl. Phys.} {\bf #1} (19#3) #2}
\def\nupps#1#2#3{{\it Nucl. Phys. (Proc. Suppl.)} {\bf #1} (19#3) #2}
\def\zp#1#2#3{{\it Z. Phys.} {\bf #1} (19#3) #2}

\def\ptp#1#2#3{{\it Prog. Theor. Phys.} {\bf #1} (19#3) #2}

\def\mpl#1#2#3{{\it Mod. Phys. Lett.} {\bf #1} (19#3) #2}
\def\pscrip#1#2#3{{\it Physica Scripta} {\bf #1} (19#3) #2}
\def\ijmp#1#2#3{{\it Intl. Jour. of Mod. Phys.} {\bf #1} (19#3) #2}
\def\abstracts#1{{
	\centering{\begin{minipage}{30pc}\tenrm\baselineskip=12pt\noindent
	\centerline{\tenrm ABSTRACT}\vspace{0.3cm}
	\parindent=0pt #1
	\end{minipage} }\par}}
\input epsf
\begin{document}
\centerline{\tenbf APPLICATIONS OF CHIRAL SYMMETRY}
\baselineskip=22pt
\vspace{0.8cm}
\centerline{\tenrm ROBERT D. PISARSKI}
\baselineskip=13pt
\centerline{\tenit Dept. of Physics, Brookhaven National Laboratory}
\baselineskip=12pt
\centerline{\tenit Upton, NY 11973, USA}
\vspace{0.9cm}
\abstracts{I discuss several topics in the applications of chiral symmetry at
nonzero temperature.  First, where does the rho go?  The answer:
up.  The restoration of chiral symmetry at a temperature $T_\chi$
implies that the $\rho$ and $a_1$ vector mesons are degenerate
in mass.  In a gauged linear sigma model the
$\rho$ mass increases with temperature,
$m_\rho(T_\chi) > m_\rho(0)$.  I conjecture that at $T_\chi$
the thermal $\rho - a_1$ peak is relatively high,
at about $\sim 1 \, GeV$, with a width approximately that at zero
temperature (up to standard kinematic factors).  The $\omega$ meson
also increases in mass, nearly degenerate with the $\rho$, but its
width grows dramatically with temperature, increasing to at least
$\sim 100 \, MeV$ by $T_\chi$.  I also stress how utterly remarkable
the principle of vector meson dominance is, when viewed from the
modern perspective of the renormalization group.
Secondly, I discuss the possible appearance of disoriented chiral
condensates from ``quenched'' heavy ion collisisons.
It appears difficult to obtain large domains of
disoriented chiral condensates in the standard two flavor model.
This leads to the last topic, which is the phase diagram for $QCD$ with
three flavors, and its proximity to the chiral critical point.
$QCD$ may be very near this chiral critical point, and one might
thereby generated large domains of disoriented chiral condensates.}
\begin{center}
{\it Based upon talks
presented at the ``Workshop on Finite Temperature QCD and
Quark-Gluon Transport Theory'', Wuhan, PRC, April, 1994}
\end{center}
\rm\baselineskip=14pt
\section{Introduction}

In this paper I review several topics in which effective models are used
to study the dynamics of chiral symmetry at nonzero temperature.
The order is somewhat jumbled, and approximately in reverse chronological
order from that in which the work was done.  Whatever else, this has the
virtue of presenting what I am most excited about (since it is most
recent) first.

\section{Vector mesons and chiral symmetry}

The spectrum of dileptons in the collisions of heavy ions at ultrarelativistic
energies provides a window into ``hot'' regions of the collison,
whereby the formation of a quark-gluon plasma might be observed.
For any effects of a quark-gluon plasma to be distinguishable from
the background of ordinary hadronic processes,
the system must last for a long period of time
at a temperature $T_\chi$, burning off the entropy of a
quark-gluon phase.  This is true for a first order transition with
a large latent heat, but
applies even if there is no true phase transition, as long as there is
a large jump in the entropy in a narrow region of temperature.
Numerical simulations appear to find such a jump in the
entropy.$^{\ref{karsch}}$

There are two distinct ways in which vector mesons at a temperature
$T_\chi$ can be affected by the quark-gluon plasma, or more generally,
by a hot hadronic plasma.  We should speak of either type of plasma,
since if there is no true phase transition the two cannot be
distinguished, even in principle.
(I speak only of nonzero
temperature, and not of a system with nonzero baryon density.
Cold, dense systems can be treated by an extension of the present methods,
but may well involve new phenomenon.)

The first is if the vector meson has a large decay width
so that its lifetime is short, less than $1 fm/c$,
and it decays inside the plasma.  Then if its effective
mass at $T_\chi$ is
different from that at zero temperature, as it surely must be,
in principle the shift in its mass, and so the corresponding peak
in the dilepton spectra, is observable.  At zero
temperature the only example which appears in the dilepton spectra
is the peak for the $\rho$ meson, with a width of $\sim 150 \, MeV$.
I suggested such a shift in the ``thermal'' $\rho$ peak some time
ago.$^{\ref{rdp1}}$

The second way in which a vector meson can be affected is if it
has a long lifetime, much greater
than $1 fm/c$, so that it decays outside of the plasma.  Then even
if its effective mass at $T_\chi$ is different from that at zero temperature,
the overwhelming bulk of decays occurs
outside of the plasma, and the peak for this vector meson in dileptons
doesn't move.  Instead, the peak shrinks, as the hot medium shakes the bound
state apart.
The $J/\Psi$ peak is suppressed in this way.$^{\ref{ms}}$

At first thought, one might expect that the $\omega$ meson is
like the $J/\Psi$ meson, since at zero temperature its width is
small relative to that of the $\rho$ meson, of order $\sim 10 \, MeV$.
In this work I argue that this expectation is wrong:
the width of the $\omega$ meson quickly grows with temperature,
and is large at $T_\chi$, at least
$\sim 100 \, MeV$.  Thus thermal $\omega$ mesons
decay inside, and not outside, the plasma, and the shift of their
masses is in principle observable.  The width of the $\rho$ meson does
not significantly increase with temperature.

The first question to settle is: where do the $\rho$ and $\omega$
mesons go?  Do their
effective masses go up, or down,
with increasing temperature?

My first guess was ``down'' --- that the effective mass decreased with
temperature.$^{\ref{rdp1}}$  If the phase transition to a quark-gluon phase
is primarily one of deconfinement, then this may
be modeled by an effective bag constant which decreases with
temperature.
The effective mass of the $\rho$ meson, like all hadronic
bound states, should then decrease with
temperature.$^{\ref{rdp1},\ref{brown},\ref{ko}}$

\begin{figure}[t]
\epsfxsize=3.2in
\centerline{\epsffile{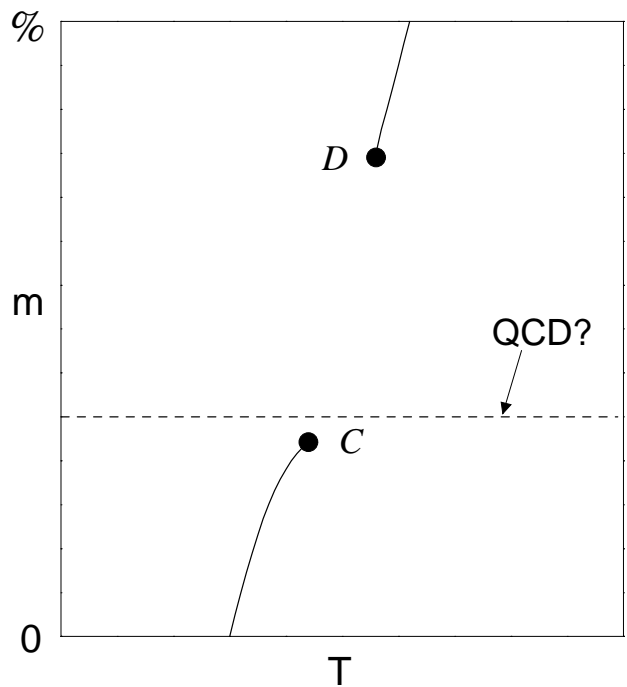}}
\caption{
The phase diagram of a gauge theory with three colors and $2+1$ flavors,
following the results of the Columbia group.
The quark masses are held in fixed ratio, so
up $=$ down quark mass, ${\cal M}$, and the strange quark mass
$= 32 {\cal M}$.
The $y$-axis is ${\cal M}$, running from $0$ to $\infty$, while the
$x$-axis is the temperature $T$.
There are two critical points,
the deconfining critical point, ${\cal M}_D$, and the chiral critical
point, ${\cal M}_C$.  As drawn, $QCD$ appears to lie above but near
the chiral critical point.}
\end{figure}

Numerical simulations of lattice gauge theory, however, demonstrate that
the phase diagram of $QCD$ with three flavors of quarks is rather
involved, and depends crucially on the values of the quark
masses.$^{\ref{karsch}}$  This material will be discussed at greater
length in sec. 3, so I just summarize the results here.
For simplicity I work with ``2+1'' flavors, holding
the up, down, and strange quarks in fixed ratio,
${\cal M}\equiv{\cal M}_{up}={\cal M}_{down}=r {\cal M}_{strange}$,
with $r\sim 1/32$.$^{\ref{rdp2}}$  At infinite
${\cal M}$ there is a deconfining phase transition of first
order, which {\it might} be --- but at least to my eyes,
today, need not be --- modeled by a decreasing bag constant.

(Perhaps it is worth elaborating on my concerns.  At infinite
${\cal M}$ we are dealing with the phase transition in the pure
glue theory.  This is first order, with masses which shift with
temperature on either side of the phase transition.  But why should
the masses decrease as the transition is approached from below?
In particular, I can certainly rule out any transition in which
the effective bag constant vanishes identically.  If this were to happen,
then by construction an {\it infinite} number of bound states would become
massless.  But there is no known critical point where an infinite
number of states become massless: for all known critical points in
$3+1$ dimensions, only a finite number of fields become massless.)

Going down from infinite ${\cal M}$ there is a line of first order
deconfining phase transitions; for three colors and $2+1$ flavors,
this line of first order deconfining phase transitions does not extend
indefinitely, but terminates in a deconfining critical point, ${\cal M_D}$.

The opposite limit of ${\cal M} =0$ is the point of chiral symmetry for
three flavors; here the chiral phase transition is expected to be of first
order.$^{\ref{pw}}$  Again, as ${\cal M}$ increases from zero there is
a line of first order chiral transitions, which for three colors
and $2+1$ flavors terminates in a chiral
critical point, ${\cal M_C}$.  Thus the crucial feature of the phase diagram
is a gap, for ${\cal M_D > M > M_C}$, in which there is no true phase
transition.  Current lattice data$^{\ref{karsch}}$ indicates that while
$QCD$ lies in this gap, ${\cal M_D > M}_{QCD} > {\cal M_C}$, ${\cal M}_{QCD}$
lies much closer to the chiral critical point than to the deconfining
critical point.  Thus $QCD$ appears to be described by a smooth
transition which is dominated by the (approximate) restoration of
chiral symmetry.  This phase diagram is illustrated in fig. (1),
and will be considered at length later.

\subsection{Chiral symmetries}

Consequently we require a model which treats the restoration of chiral
symmetry for both spin zero and spin one mesons.
Before constructing the model, it is worth reviewing how
the underlying quark fields, and so the composite meson fields constructed
from them, transform under the global chiral symmetries.

For simplicity I only consider the case of two flavors; since the
$\phi$ meson is known experimentally to be predominantly $\bar{s} s$,
this is probably adequate for treating $\rho$ and $\omega$ mesons.  It is
surely inadequate for mesons such as the $\eta$ and $\eta'$, which
are not flavor, but $SU(3)$, eigenstates.

For left and right handed quark fields $q_l$ and $q_r$,
$q_{l,r} = (1 \pm \gamma_5)q/2$, under a
global chiral symmetry transformation
in the group $SU_l(2)\times SU_r(2)\times U_a(1)$,
\begin{equation}
q_l \; \rightarrow e^{i \theta_a} \; U_l \; q_l \;\;\; , \;\;\;
q_r \; \rightarrow e^{- i \theta_a} \; U_r \; q_r \; .
\label{e1}
\end{equation}
Here $U_{l,r}$ are transformations under $SU_{l,r}(2)$, and
$e^{i \theta_a}$ that for axial $U_a(1)$.

The spin zero
mesons constructed from the quark fields include two singlet fields,
\begin{equation}
\sigma_\eta \; = \; \bar{q} \, q \;\;\; , \;\;\;
\eta \; = \; \bar{q} \, \gamma_5 \, q \; ,
\label{e2}
\end{equation}
and two isotriplet fields,
\begin{equation}
\sigma_{\vec{\pi}} \; = \; \bar{q} \, \vec{t} \, q \;\;\; , \;\;\;
\vec{\pi} \; = \; \bar{q} \, \gamma_5 \, \vec{t} \, q \; ,
\label{e3}
\end{equation}
where the $\vec{t}$ are proportional to
the Pauli matrices for $SU(2)$.  The $\eta$
and $\pi$'s are $J^P = 0^-$ states, while the $\sigma_\eta$ and
the $\sigma_{\pi}$ are $J^P = 0^+$.  The identification of the
$0^+$ states is highly controversial; the candidates in the Particle
Data Tables are the $f_0(975)$ and the $a_0(980)$.  Both of these states
have narrow widths, each less than $60 \, MeV$, and so are very
possibly some other type of state, such as a $\bar{K} K$ molecule.

The spin one mesons are constructed analogously by adding the Dirac
matrix $\gamma_\mu$ everywhere.  In terms of the plausible candidates
in the Particle Data tables these fields are the isosinglet states,
\begin{equation}
\omega^\mu \; = \; \bar{q} \, \gamma^\mu \, q \;\;\; , \;\;\;
f^\mu_1 \; = \; \bar{q} \, \gamma_5 \, \gamma^\mu \, q \; ,
\label{e4}
\end{equation}
and the isotriplet states,
\begin{equation}
\vec{\rho}^{\, \mu}
\; = \; \bar{q} \, \gamma^\mu \, \vec{t} \, q \;\;\; , \;\;\;
\vec{a}_1^{\, \mu} \; = \; \bar{q} \,
\gamma_5 \, \gamma^\mu \, \vec{t} \, q \; .
\label{e5}
\end{equation}
The $\omega(783)$, the $\rho(770)$, and the $a_1(1260)$ are all familiar;
the only unfamiliar state is the $f_1$, which I identify with
the lightest $f_1$ in the Particle Data Table, the $f_1(1285)$.
The $\omega$ and $\rho$ are $J^P = 1^-$; the $f_1$ and the $a_1$
are $J^P = 1^+$.

I would like to make a side remark.  From a study of the
$QCD$ phase diagram as a function of the current quark mass ${\cal M}$,
it is known that how far $QCD$ is from the chiral critical point depends
crucially on the value of the mass for the isosinglet $0^+$, the $\sigma_\eta$,
at zero temperature.$^{\ref{rdp2}}$
Of course in $QCD$ this is inevitably complicated
by the decay channel into two pions, by mixing with $\bar{K} K$ molecules,
and a myriad of other details.

But in the quenched approximation,
all of these details are absent, since by construction all mesons have
zero width.  Thus one can ask:

{\it In the quenched
approximation, is the mass of the $\sigma_\eta$ less than, or greater
than, the mass of the $\rho$?}

I am not suggesting that the answer
in the quenched approximation is the same as for $QCD$; it could well
differ.  I am suggesting, however, that it {\it is} a well defined question,
with an answer amenable by present day techniques.

To return to the question of chiral symmetry, under $SU_l(2) \times SU_r(2)$
transformations the scalars mix with each other as:
\begin{equation}
\sigma_\eta \; \leftrightarrow \; \vec{\pi} \;\;\; , \;\;\;
\eta \; \leftrightarrow \; \sigma_{\vec{\pi}} \; ,
\label{e6}
\end{equation}
while the vectors mix as:
\begin{equation}
\omega \; \& \; f_1, \;\;\; {\rm invariant} \;\;\; ; \;\;\;
\vec{\rho} \; \leftrightarrow \vec{a}_1 \; .
\label{e7}
\end{equation}
For the scalars, the $0^+$ and the $0^-$ states mix, while for
the vectors, while the $1^-$ $\rho$ and the $1^+$ $a_1$ mix, the
singlet states, the $\omega$ and the $f_1$, are left unchanged.
The singlets are invariant because they correspond to currents
for fermion number and axial fermion number, which are
conserved classically in the massless theory.
{}From Eqs. (\ref{e6}) and (\ref{e7}),
in a chirally symmetric phase the masses
of the $\sigma_\eta$ and the $\pi$,
the $\eta$ and the $\sigma_\pi$, and the $\rho$ and the $a_1$,
are equal to each other; there is no prediction for the masses of the
$\omega$ and the $f_1$, since they don't mix with any other states under
the chiral symmetry.

The axial $U_a(1)$ symmetry is
broken quantum mechanically.
Nevertheless, for completeness I list how the states transform
under $U_a(1)$:
\begin{equation}
\sigma_\eta \; \leftrightarrow \; \eta \;\;\; , \;\;\;
\sigma_{\vec{\pi}} \; \leftrightarrow \; \vec{\pi} \; .
\label{e8}
\end{equation}
In contrast to the scalars,
{\it all} vectors --- the $\omega$, $f_1$, $\rho$, and the $a_1$,
are invariant under $U_a(1)$ transformations.
This follows because of the conservation of $U_a(1)$ at the classical
level.

At zero temperature we know that the effects of the anomaly, and so
the quantum mechanical breaking of the axial $U_a(1)$ symmetry, are
large, because the $\eta'$ (which for two flavors is the $\eta$ meson)
is heavy.  It is reasonable to suspect that at very high temperature
the axial $U_a(1)$ symmetry is at least partially restored.
One way of seeing this is to see if the
masses of the states which mix with each other under Eq. (\ref{e8})
are approximately equal.  While this is possible for the singlet
states, the $\eta$ and the $\sigma_\eta$, it is probably much easier
for the isotriplet states, the $\vec{\pi}$ and the $\sigma_{\vec{\pi}}$.

{\it Thus on the lattice, one can indirectly look for the restoration
of axial $U_a(1)$ by measuring the mass splitting between the $\pi$'s
and the $\sigma_\pi$'s.}  This provides another motivation for measuring
the properties of the scalar particles.

In this paper
I will ignore the possible effects of the restoration of axial
$U_a(1)$, and concentrate on the restoration of $SU_l(2) \times SU_r(2)$.
For the scalars, this implies that I forget the $\eta$ and $\sigma_\pi$
fields, keeping only the $\pi$ and $\sigma_\eta$ mesons.  Since the
there is just one sigma field, the isosinglet $\sigma_\eta$,
I will refer to that simply as the $\sigma$ field.  I stress, however,
that even for two flavors there is another isotriplet of scalar fields
which in principle should be included, the $\sigma_{\vec{\pi}}$'s.
As will be discussed in sec. 3, for three flavors
there is a full nonet of sigma mesons which go along with the usual
nonet of pseudoscalar Goldstone bosons.

If we only wanted to decsribe the lightest fields at zero temperature,
which are the pions, then it would suffice to use a nonlinear sigma model.
While this is currently a very popular approach, I stress that at best
it is most awkward to describe the phase transition to a chirally
symmetric phase.  The reason is rather trivial: above the
temperature for the restoration of chiral symmetry, all particles must
fall into degenerate multiplets of $SU_l(2) \times SU_r(2)$, and possibly
$U_a(1)$ as well.  It is simply easier if we start with a theory in which
all the fields which we need to form the complete multiplets
are there to begin with, rather than having to generate them dynamically.
Of course if we don't put the full multiplet in by hand,
they will be generated dynamically.
For example, in the nonlinear
sigma model in $2+\epsilon$ dimensions, by using
the large $N$ expansion it can be shown that in the chirally symmetry
phase the $\sigma$ field is generated dynamically,
as a bound state of two pions.$^{\ref{bls}}$

Part of the prejudice against the linear sigma model, as opposed to
the nonlinear model, is due to the fact that the sigma meson is
very broad at zero temperature, from its decay mode into two
pions.  Thus the feeling is, why bother?
But even if the sigma is broad at zero temperature, inevitably it
will become narrow at high temperature.  This is trivial in a chirally
symmetric phase, since then the sigma and the pions are degenerate.
When the pion is massive at zero temperature, a narrow sigma can even
emerge before the temperature of (approximate) chiral symmetry restoration.
This happens because as the temperature is
raised, the sigma mass goes down, while the pion mass goes up;
thus the (thermal) sigma mass falls below twice the (thermal) pion
mass before the temperature of chiral symmetry restoration.$^{\ref{hk}}$

Thus the sigma meson(s), which is a broad
state, lost in the hadronic mud at zero temperature, eventually emerges
as a clean, narrow state at nonzero temperature.

\subsection{Sigma models and vector dominance}

Having fixed upon a linear sigma model, we then have to include the
vector mesons.  Here we can rely upon one of the
unjustly forgotten triumphs of the 1960's, which is the model
of vector dominance;$^{\ref{sixties} - \ref{georgi}}$
indeed, it was Sakurai's
gauge theory of the $\rho$ which lead to the standard model, and thence
to $QCD$.  I hope to emphasize, however, that whatever its historical
antecedents, that vector dominance is an amazing thing indeed.

{}From the $SU(2)$ Pauli matrices $\sigma^a$, I introduce $t^a = \sigma^a/2$
and $t^0 = {\bf 1}/2$; these matrices are normalized so that
$tr(t^a t^b) = \delta^{a b}/2$, etc.  Since we are forgetting about
the $\eta$ and the $\sigma_\pi$ fields, for the scalar fields we only
need an $SU(2) \times SU(2) \sim O(4)$ vector,
\begin{equation}
\Phi \; = \; \sigma \, t^0 \; + \; i \, \vec{\pi} \cdot \vec{t} \; .
\end{equation}
Under a $SU_l(2) \times SU_r(2)$ chiral rotation, $\Phi$ transforms as
\begin{equation}
\Phi \; \rightarrow \; \Omega_l^\dagger \; \Phi \; \Omega_r \; .
\label{e8a}
\end{equation}

For the vector fields, I introduce left and right handed fields as
$$
A_l^\mu \; = \; (\omega^\mu + f_1^\mu)
\; + \; \left( \vec{\rho}^{\, \mu}
\, + \, \vec{a}_1^{\, \mu} \right) \cdot \vec{t} \; ,
$$
\begin{equation}
A_r^\mu \; = \; (\omega^\mu - f_1^\mu)
\; + \; \left( \vec{\rho}^{\, \mu}
\, - \, \vec{a}_1^{\, \mu} \right) \cdot \vec{t} \; .
\label{e8b}
\end{equation}
The central question is how the vector fields transform under chiral
rotations.  The obvious, and certainly the most natural guess, is that
they transform as one would expect under a global chiral rotation,
which is homogeneously:
\begin{equation}
A^\mu_{l,r} \;\; {\,}^{?} {\!\!\!\!\!\!\ \rightarrow} \;\;
\Omega_{l,r}^\dagger \; A_{l,r}^\mu \; \Omega_{l,r} \; .
\label{e9}
\end{equation}
I now show that while the transformation of Eq. (\ref{e9}) is standard,
it does not lead to a model with vector dominance; the following repeats
the discussion of Gasiorowicz and Geffen.$^{\ref{sixties}}$  I introduce the
abelian field strength tensor for the left and right handed fields,
\begin{equation}
F_{l,r}^{\mu \nu}|_{not \; VDM} \; = \; \partial^\mu A_{l,r}^\nu \; - \;
\partial^\nu A_{l,r}^\mu \; ,
\label{e10}
\end{equation}
and introduce the effective lagrangian for the gauge fields,
$$
{\cal L}_{not \; VDM} \; = \;
\frac{1}{2} tr\left( \left(F_l^{\mu \nu}|_{not \; VDM} \right)^2
\; + \; \left( F_r^{\mu \nu}|_{not \; VDM} \right)^2 \right)
$$
\begin{equation}
\; + \; m^2 \; tr \left( \left( A_l^\mu \right)^2
\; + \; \left( A_r^\mu \right)^2 \right)
\end{equation}
This is a reasonable effective lagrangian for the vector fields.
It respects the global chiral symmetry of $SU_l(2) \times SU_r(2)$,
while the mass term for the gauge fields will ensure that the gauge
fields are massive.  Of course we should also add the coupling to the
scalar field $\Phi$, but let us neglect that for the time being.

The current for
${\cal L}_{not \; VDM}$ is computed
by the standard Noether construction.
For the left handed fields, say, take
$\Omega_l = exp(i {\vec \omega}_l \cdot {\vec t} )$, and compute
the infintesimal variation of the lagrangian for a spatially dependent
$\omega_l$; then
the current is the coefficient of $\partial_\mu \omega_l$.
For infintesimal $\omega_l$, in group
space the change in the vector potential
is $\delta A^\mu_l = [A^\mu_l,\omega_l]$.
Immediately we see that the mass term in
${\cal L}_{not \; VDM}$ doesn't contribute to the current,
since $\delta tr( (A_l^\mu)^2) = tr (A_l^\mu [A_l^\mu,\omega_l]) = 0$
by the cyclic property of the trace.  That doesn't mean that there
isn't a current; the abelian field strength tensor gives rise to
a perfectly fine (left handed) current,
\begin{equation}
{\vec j}^{\mu}|_{not \; VDM} \; = \;
[A^\nu_l, \partial^\mu A^\nu_l - \partial^\nu A^\mu_l] \; .
\label{e11}
\end{equation}
{}From the standpoint of a current, this is a very reasonable expression,
bilinear in the fields.  This is analogous to the current we know
for a scalar field, $\phi^* \partial_\mu \phi - \partial_\mu \phi^* \phi$,
and that for a fermion field, $\bar{\psi} \gamma^\mu \psi$.

But it's not the current required for vector dominance.  Vector dominance
is the statement that the largest coupling of hadrons to photons is
through a current {\it linear} in the appropriate vector fields, with
a coupling proportional to the mass (squared) of the vector fields.
We are not used to dealing with currents linear in the fields!

Such a current can be constructed, but at the expense of quite a leap.
After all, we are dealing with a theory which has only a {\it global}
chiral symmetry.  Vector dominance instructs us to promote this
symmetry to one which is {\it local}.  At least for myself this
was a remarkable step which I strenously resisted, to no avail.
The basic point can be understand by changing the transformation of
Eq. (\ref{e9}) to that for a local chiral rotation,
\begin{equation}
A^\mu_{l,r} \; \rightarrow \;
\frac{1}{- i {\widetilde g}} \;
\Omega_{l,r}^\dagger \; D^\mu_{l,r} \; \Omega_{l,r} \; .
\label{e12}
\end{equation}
I introduce the covariant derivatives
$D_{l,r}^\mu = \partial^\mu - i {\widetilde g} A^\mu_{l,r}$ and the
coupling constant ${\widetilde g}$.  Notice that the coupling constant
${\widetilde g}$ is that for vector dominance --- it has nothing (directly)
to do with the coupling constant of $QCD$, and is generically a large
coupling, of order one.

The amazing point is that by assuming that the vector fields transform
{\it in}homogeneously under (local) chiral rotations, we are automatically
guaranteed that the mass term gives the proper current.  This is
simply because for small $\omega_l$,
$\delta A^\mu_l = i \, \partial^\mu
\omega_l/{\widetilde g} + [A^\mu_l,\omega_l]$.
Now for the mass term, the commutator term doesn't contribute to
the current, as before.  But the inhomogeneous part of the transformation
of the gauge field then gives, trivially, a contribution to the
current as
\begin{equation}
j^\mu_l \; = \; \frac{m^2}{{\widetilde g}} \; A^\mu_l \; .
\label{e13}
\end{equation}
This is the type of expression required for vector dominance: the
current is {\it linear} in the gauge fields, with a coefficient
proportional to the mass (squared) divided by the vector coupling
constant.

Since this is the current we want, we then have to ensure that no other
terms in the lagrangian contribute to the current.  But having taken
the giant step of introducing a {\it local} chiral symmetry, the
rest is easy.  To ensure that no other terms in the effective lagrangian
contribute to the current, we require that all other terms couple to
the vector fields in a gauge invariant manner.  As long as the couplings
are gauge invariant, then, by definition the rest of
the lagrangian will be
invariant under the gauge transformation of Eq. (\ref{e12})!
Hence we introduce the nonabelian field strengths,
\begin{equation}
F_{l,r}^{\mu \nu} \; = \; \partial^\mu A_{l,r}^\nu \; - \;
\partial^\nu A_{l,r}^\mu \; - \; i {\widetilde g} \;
[A_{l,r}^\mu , A_{l,r}^\nu] \; .
\label{e14}
\end{equation}
For the scalar field $\Phi$ the appropriate covariant
derivative acts as
\begin{equation}
D^{\mu} \, \Phi \; = \; \partial^{\mu} \, \Phi \, - \,
i \, {\widetilde g} \; \left( A^{\mu}_{l} \Phi \,
- \, \Phi A^{\mu}_{r} \right) \; .
\label{e15}
\end{equation}
There is a relative minus sign between the coupling constant for
the left and right handed handed fields because from Eq. (\ref{e8a}),
$\Phi$ transforms under $\Omega_r$ and $\Omega_l^\dagger$.
Given these quantities, the effective lagrangian for the
gauged linear sigma model which respects vector dominance is then
$$
{\cal L} \; = \;
tr |D_\mu \Phi|^2 \; - \; h \, tr(\Phi) \; + \; \mu^2 \; tr|\Phi|^2
\; + \; \frac{\lambda}{2} \; (tr|\Phi|^2)^2
$$
\begin{equation}
+ \; \frac{1}{2} tr\left( \left(F_l^{\mu \nu} \right)^2
\; + \; \left(F_r^{\mu \nu} \right)^2 \right)
\; + \; m^2 \; tr \left( \left(A_l^\mu \right)^2
\, + \, \left( A_r^\mu \right)^2 \right)
\label{e16}
\end{equation}
The scalar potential is standard, with a
term linear in $\Phi$, $\sim h \, tr(\Phi)$, added to ensure that the
pions are massive.

The assumption that the spin one mesons couple to the scalar field
only through coupling which are locally gauge invariant {\it greatly}
restricts the possible couplings.  If only a global chiral symmetry
is imposed, many more terms are possible.
For example, the terms $A^\mu_l \Phi$ and $\Phi A^\mu_r$ both transform
under global chiral rotations like $\Phi$ itself, Eq. (\ref{e8a}).
Since $QCD$ preserves parity, we can require that all possible terms
be invariant under the interchange of left and right handed gauge fields.
For example, for the quartic terms,
besides $|A^\mu_l \Phi -  \Phi A^\mu_r|^2$,
the terms $|A^\mu_l \Phi +  \Phi A^\mu_r|^2$ and
$|A^\mu_l \Phi|^2 + |\Phi A^\mu_r|^2$ are also allowed under the global
chiral symmetry.

{}From the viewpoint of the renormalization group, this restriction in
possible terms is most striking.
The standard prescription of the renormalization group is the following.
Start with the terms with the lowest mass dimension, which are
usually the mass terms.  Write down all mass terms consonant
the relevant symmetries.  For terms with higher mass dimension,
which are typically cubic or quartic in the fields, and whose coupling
constants are less relevant, all possible couplings
which respect the symmetries are allowed.
For example, for a theory with an $N$ component vector $\vec{\phi}$,
if the symmetry is $O(N)$, only terms as $\vec{\phi} \cdot \vec{\phi}$
arise.  But if the mass term is only invariant under $O(N-2)$, many
more terms are possible, as long as they are $O(N-2)$ invariant.

What is so peculiar about the model of vector dominance
is that the mass term for the gauge fields is the
most relevant operator and manifestly breaks the gauge symmetry.  Yet
for all other terms --- the coupling of the gauge fields to themselves,
and the coupling of the gauge fields to the scalar fields --- one uses
this local gauge symmetry, most crucially, to greatly restrict the possible
couplings of the gauge field.  Of course what is unusual about the model
of vector dominance is that it is a {\it local}, and not simply a global,
symmetry which arises.  Somehow this must be crucial to its use and
eventual consistency.  But it seems to me as if something more interesting
than mere dusty phenomenology is afoot.

I only use the effective lagrangian of Eq. (\ref{e16}) at the tree
level.  One simplification follows immediately:
the equations of motion for the gauge field are $D_l^\mu F_l^{\mu \nu}
= m^2 A_l^\mu + j_l^\nu$, where $j_l^\mu$ is the current from the scalar
field.  To lowest order this reduces to $\partial^\mu F_l^{\mu \nu}
= m^2 A_l^\nu$, which is only consistent is the gauge field is in
Landau gauge, $\partial_\mu A_l^\mu = 0$.  Thus in perturbation
theory the propagator for the gauge field is
$(\delta^{\mu \nu} - p^\mu p^\nu/p^2)/(p^2 + m^2)$, which falls off
like $\sim 1/p^2$ at large momenta.
Of course in gauges other than Landau, the propagator behaves like
$(\delta^{\mu \nu} - p^\mu p^\nu/m^2)/(p^2 + m^2)$, and is of order
$\sim 1$ at large momenta.  So at least the ultraviolet behavior of the
gauge propagator isn't as horrible as it might be.

Of course there is still the question of how to consistently quantize
the theory.  The usual linear sigma model at least has the virtue of
perturbative renormalizability, although one may well be pushing it into
a highly nonperturbative regime.  The effective lagrangian of Eq. (\ref{e16})
does not have such a virtue, however, and it is far from clear how to
treat the standard problems of such theories:
the coupling of (unphysical) massless modes which are present in the gauge
propagator to physical modes, unitarity, lack of renormalizability at two
loop order, and so on.  It may well be possible to find some extension
of Eq. (\ref{e16}) which is (at least) renormalizable, but I haven't thought
seriously about it.

Much of the physics of vector dominance can be read off from the coupling
of the spin one fields to the scalars through the covariant derivative.
After working out the matrix algebra, one finds
$$
tr|D_\mu \Phi|^2 \; = \;
\frac{1}{2} \; \left( \partial^\mu \sigma \, + \,
\widetilde{g} \, \vec{a}^{\,\mu}_1 \cdot \vec{\pi} \right)^2
$$
\begin{equation}
\; + \; \frac{1}{2} \; \left( \partial^\mu \vec{\pi} \, + \,
\widetilde{g}(\vec{\rho}^{\, \mu} \times \vec{\pi}) \, - \,
\widetilde{g} \, \sigma \, \vec{a}_1^{\, \mu} \right)^2 \; + \;
\frac{\widetilde{g}^2}{2} \, \left( \sigma^2
+ \vec{\pi}^2 \right) \, f_1^2 \; .
\label{e17}
\end{equation}
Notice that the $\omega$ meson drops out completely:
from Eq. (\ref{e8b}),
in the vector fields $A_{l,r}$ the $\omega$ meson is proportial to the
unity matrix, and so the $\omega$ cancels in the covariant
derivative in Eq. (\ref{e15}).  There is a coupling of the $\omega$ meson,
due to the anomaly, but I will only discuss that in passing here.

The principal couplings of interest in Eq. (\ref{e17}) are the coupling
of the $\rho$ to two pions, as
$\vec{\rho}^{\, \mu} \cdot (\vec{\pi} \times \partial^\mu \vec{\pi})$,
and the coupling of the $a_1$ to the $\sigma$ and $\vec{\pi}$,
$\sigma \, \vec{a}_1^{\, \mu} \cdot \vec{\pi}$.  Both of these couplings are
proportional to the coupling of vector dominance, $\widetilde{g}$,
and are presumably responsible for the principal decays of these
particles.

To obtain more realistic expressions, I assume that the potential
of the scalars is such that a vacuum expectation value for the sigma
field develops, $\mu^2 = \mu_0^2 < 0$.  Because of the background magnetic
field $\sim h$, the vacuum expectation value of $\phi$ must be along
the $\sigma$ direction.
After shifting $\sigma \rightarrow \sigma_0 +
\sigma$, however, a complication arises which is special to the gauged
sigma model: there is a cross term
between the $a_1$ and the pion, as
$\widetilde{g} \, \vec{a}_1^{\, \mu}
\cdot \partial^\mu \vec{\pi}$.  This cross
term must be eliminated by a shift in the $a_1$ field,
\begin{equation}
\vec{a}_1^{\, \mu} \; \rightarrow \;
\vec{a}_1^{\, \mu} \; - \; \frac{g \sigma_0}{m^2 \, + \, (g \sigma_0)^2}
\; \partial^\mu \vec{\pi} \; .
\label{e18}
\end{equation}
One can show that the equations of motion for the shifted $a_1$ field
are such that the shifted (and not the unshifted) $a_1$ field
should be in Landau gauge.

After this shift, for the vector fields the mass term
written in Eq. (\ref{e16}), $tr(A_{l,r}^2)$, gives
\begin{equation}
m^2_\rho \; = \; m^2_\omega  \; = \; m^2 \;\;\; , \;\;\;
m^2_{a_1} \; = \; m^2_{f_1} \; = \; m^2 + (\widetilde{g} \sigma_0)^2 \; .
\label{e19}
\end{equation}
This is in good accord with experiment, where the $\rho$ and the $\omega$,
as well as the $a_1$ and the $f_1$, are very nearly degenerate.  Yet
I find it mystifying.  Besides the mass term
$tr(A_{l,r}^2)$, there is a second set of mass terms
which are invariant
under global $SU_l(2)\times SU_r(2)$ rotations,
$(tr(A_{l,r}))^2$ .  These new
mass terms only contribute to the masses of the $\omega$ and the $f_1$,
but experimentally they do not seem to be there: the masses of the
$\omega$ and the $\rho$, and the masses of the $a_1$ and the $f_1$, are
each within $2 \%$ of each other!
Perhaps my understanding of vector dominance is faulty, but there appears
to be a further principle at work,
something that tells us that such flavor singlet
mass terms are very small.  Of course such singlet mesons can only be treated
realistically within the context of a full, three flavor model, but I
believe that the paradox will remain.

After shifting the $a_1$ field as in Eq. (\ref{e18}), the kinetic term for
the $\pi$ becomes
\begin{equation}
\frac{1}{2} \;
\left( \frac{m^2}{m^2 + (\widetilde{g} \sigma_0)^2}\right)
\left( \partial^\mu \vec{\pi} \right)^2 \; .
\label{e20}
\end{equation}
This demonstrates what is known as the ``partial'' Higgs effect.
In the limit that the mass of the vector meson becomes very heavy,
$m \rightarrow \infty$, all effects of the vector mesons should
decouple, and the pions (sigmas, etc.) are unaffected.  The opposite
limit, when the mass of the vector meson vanishes, $m \rightarrow 0$,
is also familiar.  From Eq. (\ref{e20}), the kinetic term for the pions
vanishes as $m \rightarrow 0$, but this is simply because in this
limit there is a true gauge symmetry and a true Higgs effect, with
the pions turning into the longitudinal components of the $a_1$'s.
For finite but nonzero $m$, what happens is that ratio
$\sqrt{m^2 + (\widetilde{g} \sigma_0)^2}/m = m_{a_1}/m_\rho$ enters.
Experimentally this is a significant number, $\sim 1.6$.

In this vein, I should also mention what is known as the KSRF
relation.$^{\ref{sixties}}$  This relation predicts that the ratio
of the $a_1$ to $\rho$ mass is fixed, $= \sqrt{2}$.  Within the general
context of gauged linear sigma models, there is no reason why the
KSRF relation should be satisfied; see, for example, the comments of
Lee and Nieh.$^{\ref{sixties}}$  One could argue that the actual
value of $m_{a_1}/m_\rho \sim 1.6$ is not too far from
$\sqrt{2} \sim 1.414...$, but I see no particular virtue in adhering
to the KSRF relation.  Instead, I use the $m_{a_1}/m_\rho$ ratio to
fix the value of the vector meson coupling $\widetilde{g}$.

After rescaling $\pi \rightarrow (m_{a_1}/m_\rho)\pi$, the physical
quantities of interest are:
\begin{equation}
m_\pi^2 \; = \; \left( \frac{m_{a_1}}{m_\rho} \right)^2
\; \frac{h}{\sigma_0} \;\;\; , \;\;\;
m^2_\sigma \; = \; \frac{h}{\sigma_0} \, + \, 2 \, \lambda \, \sigma_0^2
\;\;\; , \;\;\;
f_\pi \; = \; \frac{m_\rho}{m_{a_1 }} \, \sigma_0 \; .
\label{e21}
\end{equation}
Fitting the
above expressions with $m_\pi = 137 \, MeV$, $m_\sigma = 600 \, MeV$,
$m_\rho = 770 \, MeV$, and $m_{a_1} = 1260 \, MeV$, the results are
$$
\sigma_0 \; = \; 152 \, MeV \;\;\; , \;\;\;
h \; = \; (102 \, MeV)^3 \;\;\; , \;\;\;
$$
\begin{equation}
\mu_0^2 \; = \; - \,(588 \, MeV)^2 \;\;\; , \;\;\;
\lambda \; = \; 7.6 \; .
\label{e22}
\end{equation}
For comparison, if we send the mass of the vector mesons to infinity,
and denote the analogous quantities by a superscript $nv$ for ``no vectors'',
with the same masses for the $\pi$ and the $\sigma$ I find
$$
\sigma^{nv}_0 \; = \; 93 \, MeV \;\;\; , \;\;\;
h^{nv} \; = \; (120 \, MeV)^3 \;\;\; , \;\;\;
$$
\begin{equation}
(\mu_0^{nv})^2 \; = \; - \,(567 \, MeV)^2 \;\;\; , \;\;\;
\lambda^{nv} \; = \; 19.7 \; .
\label{e23}
\end{equation}

{}From Eq. (\ref{e21}),
the ``partial'' Higgs effect tends to increase the pion
mass and to decrease the pion decay constant $f_\pi$.
This description
is somewhat misleading, because it is phrased in terms of the quantities
$h$, $\sigma_0$, and $\mu^2_0$,
which are not directly physical; the physical
quantities are the masses and $f_\pi$.
One quantity which is physical is the value of the scalar self coupling,
$\lambda$; that is a dimensionless number, which tells one how close
the linear model is to the nonlinear model, the nonlinear model being
recovered as $\lambda \rightarrow \infty$.
I find it notable that adding the
vectors changes this coupling from $\lambda^{nv} = 19.7$ to a much
smaller value, $\lambda = 7.6$.  Presumably with the $\rho$ and $a_1$
vector mesons, some of the scalar interactions, which otherwise arise
solely from their self interaction, $\lambda (|\Phi|^2)^2$, can be
made up by exchange of vector mesons, giving a smaller value of $\lambda$.

In conclude this section by discussing how vector meson dominance works
in the limit of a large number of colors, $N_c \rightarrow \infty$;
if the nonabelian coupling of the $SU(N_c)$ gauge theory is $g^2$,
$g^2 N_c \equiv \alpha_s$ is held fixed and of order
one as $N_c \rightarrow \infty$.$^{\ref{largeN}}$
Consider, in particular, the spectrum of dileptons, which is given
by the imaginary part of the two point function
of hadronic currents, $\langle j^\mu(x) j^\nu(0) \rangle$.  In perturbation
theory this graph starts with a free quark loop, which is of order
$N_c$ for quarks in the fundamental representation of
$SU(N_c)$, plus an infinite series of corrections in $\alpha_s$.
Now at large $N_c$ it is known that all color singlet currents are
saturated by hadronic states, where mesonic masses are of order
one.  This works fine for the vector channel
considered here if the coupling of $j^\mu$ to the $\rho$ is of order
$\sqrt{N_c}$, which is standard at large $N_c$.

This can also be seen from vector dominance, {\it assuming} that vector
dominance holds in the large $N_c$ limit.  From Eq. (\ref{e13}),
$j^\mu = m^2 \, A^\mu/{\widetilde g} $; $m = m_\rho$ is of order one,
as is the vector field $A^\mu$, since it is just another mesonic field.
The coupling of vector meson dominance, however, is a trilinear coupling
between three mesons.  This is typically vanishes
at large $N_c$ as ${\widetilde g} \sim 1/\sqrt{N_c}$, so
Eq. (\ref{e13}) becomes $j^\mu \sim \sqrt{N_c} \, A^\mu$, and agrees
with the counting of quark loops.

All of this is totally standard.  Let us then push on and
consider corrections to the
large $N_c$ limit.  The coupling of the $\rho$ to two pions is again
a trilinear meson coupling,
of order ${\widetilde g} \sim 1/\sqrt{N_c}$, so the
width of the $\rho$ meson is of order ${\widetilde g}^2 \sim 1/N_c$.
Thus while the height
of the $\rho$ peak is very large, of order $N_c$, it is very narrow,
of order $1/N_c$, so the total area under the $\rho$ peak is of order
one.  At order one there are other contributions to dilepton
production, such as from pion pairs.
Since the coupling of hadrons to photons is electromagnetic,
and pions don't interact at large $N_c$, the pion contribution
to dileptons is just given by the imaginary part of the {\it one}
loop diagram of pions.
Since both the electromagnetic coupling of the pions and their
number are of order one, so is the total pion contribution.
Thus the large $N_c$ limit is
consistent with vector meson dominance as follows:
vector mesons, such as the $\rho$, $\omega$, {\it etc.} give tall
but narrow peaks, under which is a smooth continuum for $\pi^+ \pi^-$,
$K^+ K^-$ pairs, and so on.
The total area under the vector meson peaks, and the continuum from
scalar meson pairs, is the same, of order one.

The above counting just shows that the large $N_c$ limit is compatible
with vector meson dominance; it does {\it not} show that in fact
vector meson dominance emerges at large $N_c$.
Nevertheless, it is important to see that there is nothing
inconsistent between vector meson domainance and the limit of large $N_c$.

\subsection{Vector mesons at nonzero temperature}

With all of this introduction, the extension to nonzero temperature
is relatively straightforward.  I deal exclusively with mean field
theory, since at least that gives a well defined approximation.  I am
not claiming that it is a good approximation, simply that it is
well defined.

Consider the mass terms in the effective lagrangian of Eq. (\ref{e16}),
\begin{equation}
{\cal L}_{mass} \; = \;
\mu^2 \; tr|\Phi|^2
\; + \; m^2 \; tr \left( \left(A_l^\mu \right)^2
\, + \, \left( A_r^\mu \right)^2 \right)
\label{e24}
\end{equation}
Diagramatically, {\it both} masses increase
with increasing temperature.  For example, from the scalar self interactions,
a term as $\sim \lambda T^2 tr|\Phi|^2$ is generated from the interactions
with a pion loop; similarly, the interactions of vector mesons with
a pion loop also produce a term $\sim {\widetilde g}^2 T^2
tr(A_l^2 + A_r^2)$.  {\it A priori}, I do not see any reason
why the contribution of a pion loop to one term should be significantly
larger, or smaller, than the other.

This allows me to make what is mathematically trivial, but physically
profound, observation: that the effects of thermal fluctuations
are {\it always} to cause {\it both} $\mu^2$
and $m^2$ to increase with
increasing temperature.  The details of their relative increase
surely depends on details of the model --- the number of flavors, colors,
the values of the self couplings, {\it etc.}  But in {\it any} gauged
sigma model, both mass terms go one way and only one way
with increasing temperature, and that is {\it up}.

The expressions for the masses of the vector mesons given in Eq. (\ref{e19})
remain valid as $m^2$ and $\mu^2$ (and so $\sigma_0^2$) change.
Thus a gauged sigma model predicts that the mass of the $\rho$ increases
{\it monotonically} with temperature; similarly, that the masses of
the $\omega$ and the $\phi$ also increase.
Because of the constraints of chiral symmetry, the masses of the
other vector mesons, the $a_1$ and the $f_1$ (and their
strange partners) behave in a more complicated
fashion: while $m^2$ increases, $\mu^2$ becomes less negative,
so $\sigma_0$ decreases.  Thus the $m_{a_1}$ and $m_{f_1}$ first decrease
with temperature, until they become (approximately) degenerate with
the $\rho - \omega$; then the masses of all states continue to increase
with temperature.

The above assumes, implicitly,
that there are no singlet masses for the vector fields,
$(tr(A_{l,r}))^2$.  Such singlet mass will split
the $\omega$ and $f_1$ apart from
the $\rho$ and $a_1$.  Even if such a term is not present at
zero temperature --- because
of experimental constraints on the known masses of the $\rho$ and $\omega$ ---
unless there is a symmetry reason why such a term cannot arise, it will
be generated at nonzero temperature.
If such a term is generated, the degeneracy between the $\omega$ and the
$\rho$ will be lifted, but both masses will still increase with temperature.
In any case, as we shall see, all states becomes
broad at nonzero temperature, so a relatively small shift in the real
part of the masses may not be significant.

The prediction that the mass of the $\rho$ (and those of the
$\omega$ and $\phi$) increase with temperature is not standard lore.
It is a prediction of a gauged sigma model, although
the generality of the remark does not seem to have been stressed in
previous studies.$^{\ref{lsT}}$  Different models, such as those of Brown
and Rho$^{\ref{brown},\ref{ko}}$
predict the opposite: that the mass of the $\rho$
decreases as the temperature goes up.  In principle this could be studied
by means of other phenomenological approaches, such as sum rules.  Here
one runs into a problem: extending sum rules to nonzero
temperature is not free of ambiguity.  Thus while some papers find
that the mass of the $\rho$ decreases with increasing
temperature,$^{\ref{srT}}$
others find that it increases.$^{\ref{osrT}}$  Besides my obvious prejudice
in the matter, it is worth noting that the studies which claim to find
that the $\rho$ mass increase with temperature find that it does so
in a manner consistent with the restoration of chiral symmetry.
Of course chiral symmetry alone is not enough to claim that the $\rho$
mass goes up with temperature; all chiral symmetry tells you is that the
$\rho$ and $a_1$ should be approximately degenerate.

One approach which differs from a gauged linear sigma model,
and yet incorporates
chiral symmetry in a crucial way, is the study of Nambu-Jona-Lasino (NJL)
models.$^{\ref{vmd}}$  If the predictions of the gauged linear sigma model
have any generality whatsoever, they should hold for a NJL.  There is
a technical caveat: what is usually studied are (chirally) symmetric
NJL models with the simplest possible interaction,
$(\bar{\psi} \vec{t} \psi)^2 + (\bar{\psi} \vec{t} \gamma_5 \psi)^2 $.
This type of interaction is certainly adequate for treating the scalar
and pseudoscalar particles, since undoing the quartic interactions by
means of auxiliary fields with automatically generate effective fields
for these particles.  To study vector and axial vector particles, however,
I suggest that it is {\it crucial} to include interactions such as
$(\bar{\psi} \vec{t} \gamma^\mu \psi)^2
+ (\bar{\psi} \vec{t} \gamma_5 \gamma^\mu \psi)^2 $, in what is termed
the extended NJL model.

With this cautionary aside,
there is a unique prediction: in any extended NJL model,
the mass of the $\rho$ should increase, monotonically, with temperature.
If not, there is something seriously wrong with a gauged linear sigma
model, or with vector meson dominance.

In principle, the masses of the vector mesons could also be studied by
numerical simulations of lattice gauge theory.  This is not elementary
to date all studies have
concentrated on the (manageable) task of computing static correlation
lengths.  While the $\pi$ and $\sigma$ channels are special, in {\it all}
other channels one finds that two quark states have a static correlation
length $\sim 1/(2 \pi T)$, three quark states one of $\sim 1/(3 \pi T)$,
and so on.  The usual interpretation is that this represents the propagation
of free quarks, dressed by interactions.  I suggest that if the deconfining
transition is far afield, then what one is seeing really is the propagation
of mesons (and baryons) in a hot hadronic gas.  To definitively answer
the problem, however, it is necessary to look not at static correlation
lengths, but at the true poles in the appropriate (effective) propagators.
These are defined in real time, by the analytic continuation of the
spectral densities from euclidean momenta.  Thus it is necessary to measure
not just the static correlation lengths, at $p^0 = 0$, but the correlation
lengths for $p^0 = \pm 2 \pi T$, $\pm 4 \pi T$, {\it etc.}, and then
fourier transform.  This is a very difficult problem.

At this point I note an alternate scenario
by Georgi.$^{\ref{georgi}}$  He works in a nonlinear model, with
the term for the $\rho$ mass squared
proportional to a dimensionless coupling constant times $f_\pi^2$.
He then argues that as $f_\pi \rightarrow 0$, $m_\rho \rightarrow 0$.
In terms of the linear sigma model, it is not apparent to me why
the mass term for the $\rho$ should involve $f_\pi$ at all; why is it
simply not some other dimensional parameter?
One can rephrase the argument in terms of the linear sigma model:
Georgi's term corresponds to assuming that there is
no bare mass term for the $\rho$, such as
$m^2 tr(A_l^2 + A_r^2)$, but only a term as
$tr(\Phi^\dagger \Phi) tr(A_l^2 + A_r^2)$,
which would have a dimensionless coupling constant.
In the linear sigma model
this is odd: why should there be no bare mass term
for the vector mesons, yet one induced by spontaneous symmetry breaking?
After all, the bare mass term is more relevant than Georgi's term.
Leaving prejudice aside, I appeal to the lattice data.
Admittedly, while static screening lengths are not automatically the
(inverse) pole masses, it is true that if the $\rho$ were massless
at $T_\chi$, then its static screening length would diverge, like
that for the $\sigma$ and the $\pi$.  This
is not seen in present day lattice simulations: only the $\sigma$ and
the $\pi$ have divergent static screening lengths (at least for two
flavors, where the transition appears to be of second order).

Returning to the matter at hand, one of the drawbacks of the gauged
sigma model is that it does not predict how much $\mu^2$ and $m^2$
increase with temperature.  I thus adopt what is, most honestly, a wild
guess.  As the temperature increases, the mass of the $\rho$ goes up,
while that of the $a_1$ goes down, until they meet (approximately),
after which they both increase.  I define the temperature of chiral
symmetry restoration $T_\chi$ as that for which the two masses are
approximately equal, and {\it assume} that they obey the rule
that their mass at $T_\chi$ is the arithmetic mean of their masses at
zero temperature,
\begin{equation}
m_\rho (T_\chi) \approx m_{a_1}(T_\chi) \sim
\frac{1}{2} \; \left( m_\rho(0) \, + \, m_{a_1}(0) \right) \;
= \; 1 \, GeV \; .
\label{e25}
\end{equation}
I emphasize that this is only a guess; at least it is easy
to remember.  It probably is an upper bound
on how much the $\rho$ mass can increase with temperature.  For example,
Song$^{\ref{lsT}}$ has done calculations in a
gauged {\it non}linear sigma model,
and finds that with different parametrizations, the mass of the $\rho$
can even stay (almost) constant with temperature, or increase as much
as in Eq. (\ref{e25}).  The former seems extremely unlikely
to me; again, the pion loops that drive $\mu^2$ up also contribute
to driving $m^2$ up.  Nevertheless, how much the $\rho$ mass increases
with temperature is probably only something which can be settled by
lattice simulations.

\begin{figure}[t]
\centerline{\epsffile{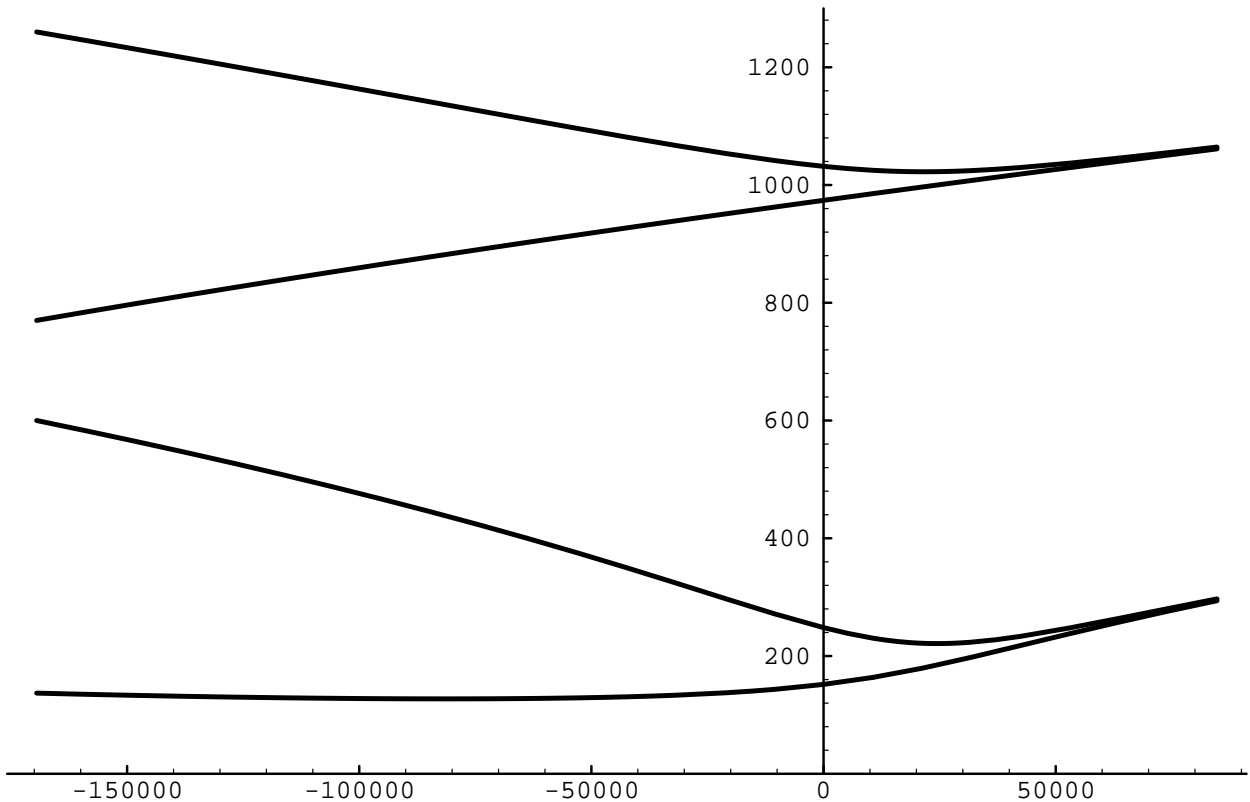}}
\caption{
Masses of the vector and scalar particles versus the mass parameter
$\mu^2$ in a two flavor, gauged linear sigma
model.  The masses are plotted against the $y$-axis in $MeV$;
from top to bottom, these are the masses of the $a_1$,
$\rho$, $\sigma$, and $\pi$, respectively.
$\mu^2$ is
along the $x$-axis: the values at zero temperature are at the extreme
left, with increasing $\mu^2$ equivalent to increasing temperature squared.
}
\end{figure}

The virtue of Eq. (\ref{e25}) is that it gives a unique parametrization
of how masses change with temperature; I find that if
$\mu_0^2 = - (588 \, MeV)^2$ and $m_0^2 = (770 \, MeV)^2$ are the
values at zero temperature, then Eq. (\ref{e25}) is satisfied for
\begin{equation}
m^2 \; = \; m^2_0 \; + \; .6 \, \frac{(\mu^2 - \mu_0^2)}{\mu_0^2} \; .
\label{e26}
\end{equation}
One can then compute the properties of the model as a function of a
single parameter, $\mu^2$; all effects of increasing temperature are
then modeled by increasing $\mu^2$.  Again, I stress that Eqs. (\ref{e25})
and (\ref{e26}) are only guesses.

In fig. (2) I give the values of the masses of the $\pi$, $\sigma$,
$\rho$, and $a_1$ as a function of increasing $\mu^2$ ($\sim$ temperature
squared).  There is no prediction whatsoever for $T_\chi$, since
one can only see that there is first approximate chiral symmetry
restoration for $\mu_\chi^2 \sim + (150 \, MeV)^2$.
AT $T_\chi$, $m_\sigma \sim m_\pi \sim 200 \, MeV$; by assumption,
$m_\rho \sim m_{a_1} \sim 1 \, GeV$.  In fig. (3) I show
$f_\pi$ versus $\mu^2$; the relevant observation is that at $\mu_\chi^2$,
$f_\pi$ has decreased to about a third of its value at zero temperature.
This decrease in $f_\pi$ is elementary: for vanishing pion mass $f_\pi
\sim \sigma$, and vanishes at the temperature for chiral restoration
(if the transition is second order).

\begin{figure}[t]
\centerline{\epsffile{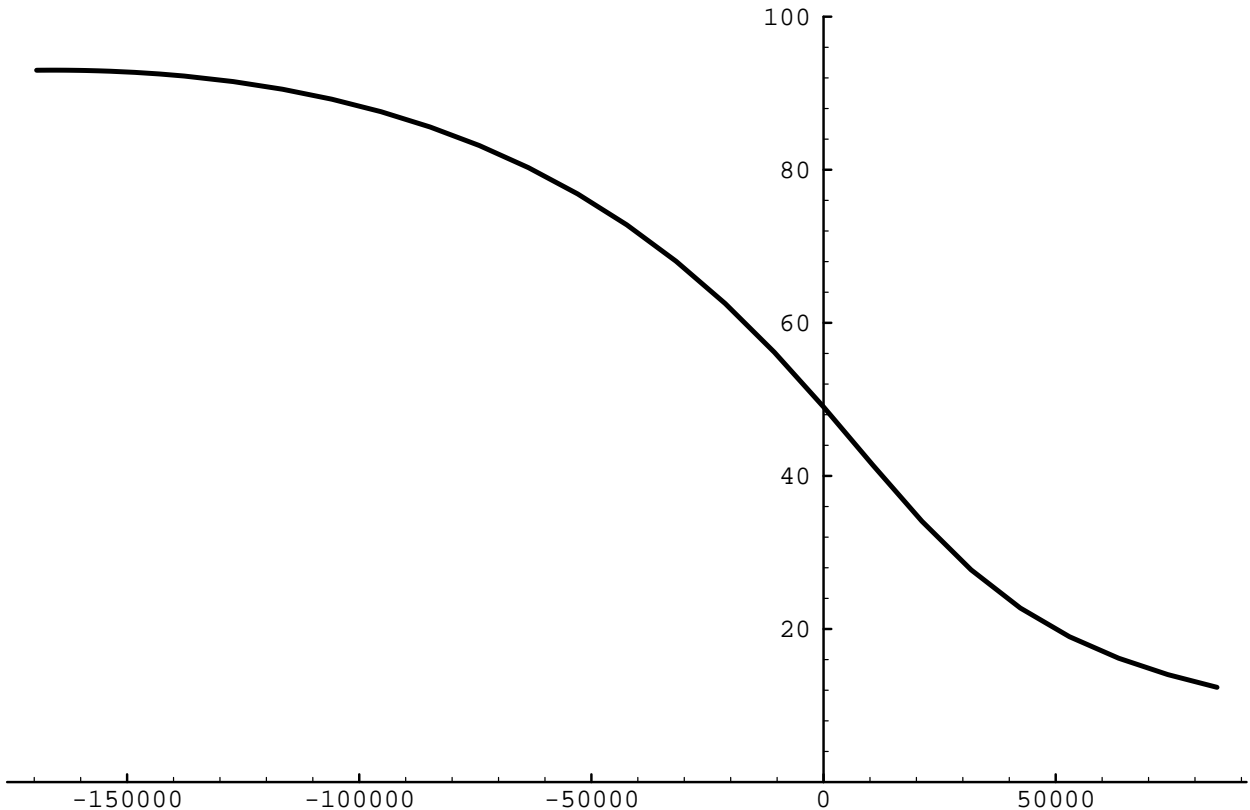}}
\caption{
The value of the pion decay constant, $f_\pi$, versus the mass parameter
$\mu^2$ in a two flavor, gauged linear sigma model.  The $y$-axis
is the value of $f_\pi$ in $MeV$, while
$\mu^2$ is
along the $x$-axis: the values at zero temperature are at the extreme
left, with increasing $\mu^2$ equivalent to increasing temperature squared.
}
\end{figure}

We can use the effective model to make more detailed predictions about
the widths of the states at $T_\chi$.  At present I give only a very
crude estimate, with careful estimates given later.$^{\ref{rdp3}}$
There is certainly new kinematics
which opens up: at zero temperature I assume that
$m_\sigma = 600 \, MeV$, so the $\rho$ can't decay into $\sigma$'s plus
$\pi$'s.  That changes at $T_\chi$, since then $m_\sigma \sim m_\pi$,
and furthermore the $\rho$ is heavier.

\begin{figure}[t]
\centerline{\epsffile{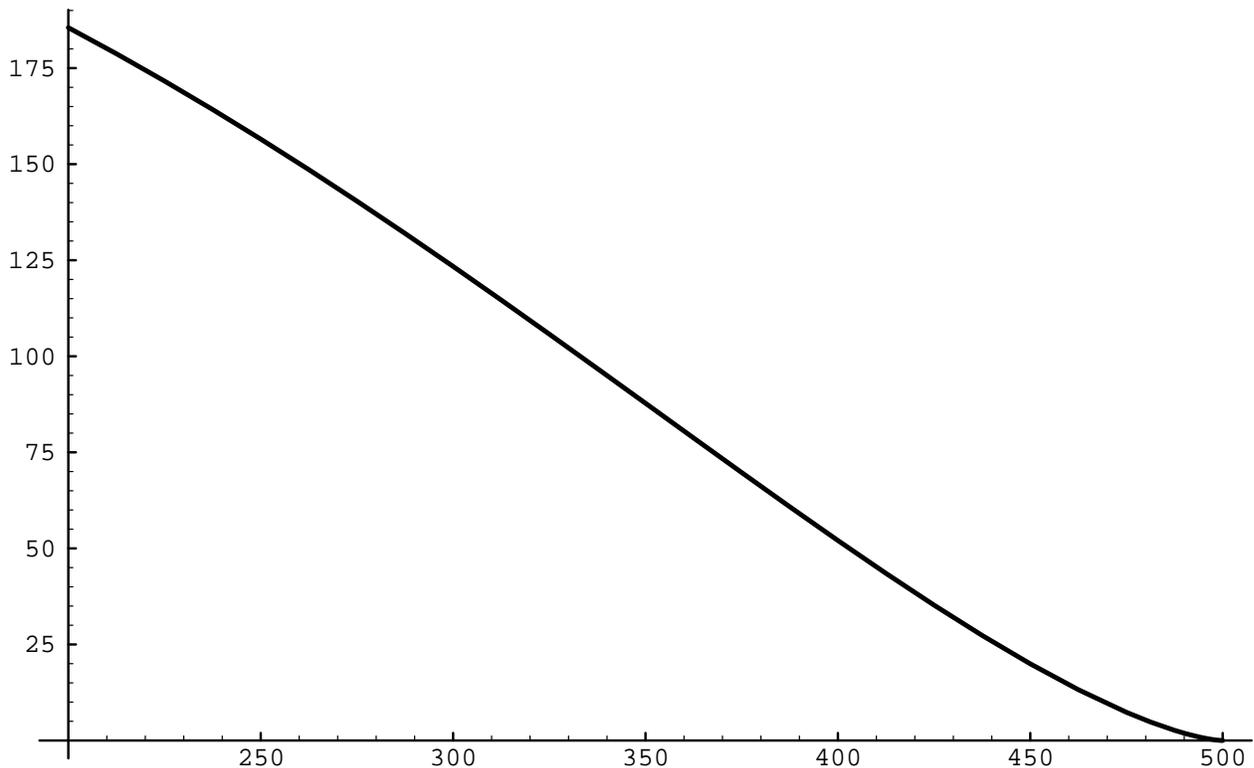}}
\caption{
The width of the $\rho$, under the extreme assumptions discussed in
the text.  Essentially, the formula for the
width is assumed to be the same as at zero
temperature, with the only difference from the mass shifts of the
$\rho$ and $\pi$.  The $y$-axis
is $\Gamma_\rho(T_\chi)$ in $MeV$, the $x$-axis the value of
$m_\pi(T_\chi)$ in $MeV$.}
\end{figure}

The decay modes of the thermal $\rho$ can be read off from the effective
lagrangian of Eq. (\ref{e17}), but the allowed channels follow from
general principles.  Even if it is
possible kinematically, the $\rho$ doesn't decay
into two $\sigma$'s because of isospin symmetry;
also, $\rho$ doesn't decay into
$\sigma \pi$ because of $G$-parity.  Of course at nonzero
temperature $\rho \rightarrow \pi \pi$,
as at zero temperature.  By similar arguments one can see that the only
new three body decay mode is $\rho \rightarrow \pi \pi \sigma$; this
respects both isospin and $G$-parity, and with the assumed masses
has nonzero phase space.
However, I do not believe that this three body decay gives
a significant contribution; the coupling constant for
$\rho \rightarrow \pi \pi$ is $\sim \widetilde{g}$; from Eq. (\ref{e17}),
that for $\rho \rightarrow \pi \pi \sigma$
is $\widetilde{g}$ times
$\widetilde{g}^2 f_\pi m_\sigma/m_\rho^2$, or $\sim \widetilde{g}/3$.
Thus the amplitude for $\rho \rightarrow \pi \pi \sigma$ is about
10\% that that for $\rho \rightarrow \pi \pi$, even without the further
restrictions which arise kinematically from a three body, versus a two body,
decay.  For this reasons I neglect the three body decays of the thermal
$\rho$.  For the remaining mode of $\rho \rightarrow \pi \pi$, I make
life even easier for myself by neglecting thermal effects entirely,
computing the decay as if it were at zero temperature.  This is valid
because the pions are so energetic; if $n(E) = 1/(exp(E/T) -1 )$ is
the Bose-Einstein distribution function for the pions,
at $E \sim 500 \, MeV$ and $T \sim 150 \, MeV$, $2 n(E/T) \sim .07$.
Consequently,
\begin{equation}
\Gamma_\rho (T_\chi) \; \sim \;
\frac{\left(m_\rho(T_\chi)^2
\, - \, 4 m_\pi^2(T_\chi) \right)^{3/2}}{m^2_\rho(T_\chi)} \;
\Gamma_\rho(0) \; \sim \; 175 \, MeV \; .
\label{e27}
\end{equation}
As the mass of the $\pi$ goes up, the decay width of the $\rho$ goes
down, to about $100 \, MeV$ for $m_\rho \sim 1 \, GeV$ and $m_\pi \sim
300 \, MeV$.  Given the crudeness of the model, I conclude that the
width of the thermal $\rho$ at $1 GeV$ is approximately equal to its
width at zero temperature.  In fig. (4) I present a graph of
$\Gamma_\rho(T_\chi)$ for $m_\rho = 1 \, GeV$, and different values of
$m_\pi$, under the above assumptions.  Due to phase space, the decay
width decreases from that in Eq. (\ref{e27}).

Implicitly I am assuming that the coupling constant for vector
meson dominance, ${\widetilde g}$, does not change with temperature.
This is probably wrong, but consideration of the one loop $\beta$-functions
suggests that ${\widetilde g}$ should decrease with temperature near
a critical point, making the width even smaller.

Computing the width of the $\omega$ meson is more difficult.  A proper
analysis requires an understanding of the Wess-Zumino-Witten
term,$^{\ref{wzw}}$
which I defer for another day.$^{\ref{rdp3}}$  For now I make a very
simple assumption, that the decay of the $\omega$ is dominated by
what is known as the Gell-Mann Sharp Wagner (GSW) mechanism.  This
is the statement that the decay $\omega \rightarrow 3 \, \pi$ proceeds
by $\omega \rightarrow \rho \, \pi$, with the virtual $\rho$ decaying
as $\rho \rightarrow \pi \pi$.  The amplitude
for the first process, $\omega \rightarrow
\rho \, \pi$, is dominated by the anomaly, and proportional to $1/f_\pi$;
the amplitude for the second process is proportional to ${\widetilde g}$
times kinematic factors.  At present$^{\ref{rdp3}}$ I ignore all of these
details to concentrate simply on the change in $f_\pi$ with temperature;
since $f_\pi(T_\chi) \sim f_\pi(0)/3$ and $\Gamma_\omega(0) \sim 10 \, MeV$,
\begin{equation}
\Gamma_\omega(T_\chi) \; \sim \;
\left(\frac{f_\pi(0)}{f_\pi(T_\chi)} \right)^2
\Gamma_\omega(0)  \; \sim \; 100 \, MeV \; .
\label{e28}
\end{equation}
This estimate is very crude, and clearly requires a more
careful analysis.$^{\ref{rdp3}}$
The basic point, however, is bound to be correct:
as chiral symmetry is (approximately) restored, $f_\pi$ decreases,
leading to a much broader $\omega$ meson than at zero temperature.
Kinematic factors will only help, as the $\omega$ moves up in mass,
so the allowed phase space increases as well.

\begin{figure}[t]
\centerline{\epsffile{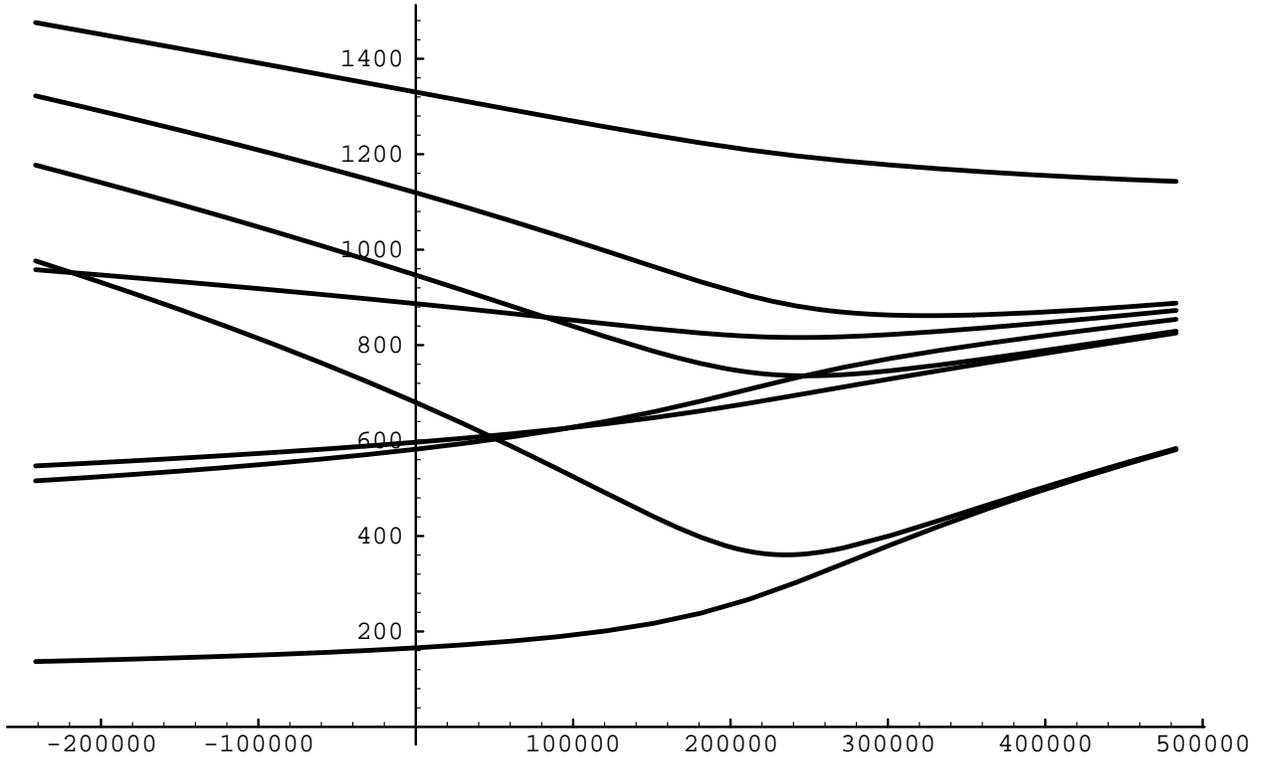}}
\caption{
Masses of the scalar particles versus the mass parameter
$\mu^2$ in a three flavor linear sigma model which is {\it not}
gauged.  The masses are plotted against the $y$-axis in $MeV$;
$\mu^2$ is
along the $x$-axis: the values at zero temperature are at the extreme
left, with increasing $\mu^2$ equivalent to increasing temperature squared.
Notice that unlike the two flavor model, for three flavors I take
$m_{\sigma_{\eta '}}(0) = 975 \, MeV$.  The assumptions which produce
this plot are described in section (3.2.1).  The point of note here is
to see that the kaon increases in mass, to approximately $650 \, MeV$.}
\end{figure}

Lastly, although the model discussed does not incorporate it, it is
reasonable to ask about the $\phi$ meson.  The chiral partner of the
$\phi(1020)$ is one of the two $f_1$ states, either the $f_1(1420)$
or the $f_1(1510)$.  There is some controversy concerning the $f_1(1420)$,
so I take the $f_1(1420)$ as some sort of special quark state, and assume
that the chiral partner of the $\phi$ is the $f_1(1510)$.  Using the same
kind of simple minded prediction for the mass at $T_\chi$ as in
Eq. (\ref{e25}), I suggest that
\begin{equation}
m_\phi (T_\chi) \approx m_{f_1}(T_\chi) \sim
\frac{1}{2} \; \left( m_\phi(0) \, + \, m_{f_1(1510)}(0) \right)
\; = \; 1250 \, MeV \; .
\label{e29}
\end{equation}
Even with this generous increase in the mass of the thermal $\phi$,
the thermal $K$ mass is like that of the $\pi$,
and also increases, to at least
$650 \, MeV$; see, for example,
fig. (5).$^{\ref{rdp3}}$  With these numbers, the $\phi$ does not have
phase space to decay as $\phi \rightarrow K \, K$ (except for thermal
processes, which should be small).  Even if the masses were to change
so that $\phi \rightarrow K \, K$ were allowed, it is very unlikely
for the width of the $\phi$ to change so much that it decays inside the
plasma.

In summary, for the thermal vector mesons at the temperature of
chiral symmetry restoration, $T_\chi$, the width of the $\rho$ remains
about that at zero temperature, and decays inside the plasma; the
$\omega$ becomes much broader, so much so that most thermal $\omega$'s
decay inside the plasma; the $\phi$ remains narrow, decaying primarily
outside of the plasma.  I do not expect a more careful analysis$^{\ref{rdp3}}$
to significantly alter these conclusions.

\section{Disoriented Chiral Condensates}

If an extraordinarily large number of pions are produced in a
hadronic collision, it is natural to think that these unusual
events might proceed by the coherent decay
of a (semi-) classical pion field.$^{\ref{anselm},\ref{ar}}$
There might be a very distinctive signature of such a coherent decay:
a given classical pion field points in some direction (or directions)
in isospin space, so coherent production is necessarily one in which
isospin symmetry would appear to be, at least locally, badly violated.
That is, most of the pions from a given region would be primarily
charged, or conversely, primarily neutral.  (Of course in the total
collision, isospin will always, in the end, average out.)
Such behavior {\it may} have been observed in Centauro events in cosmic ray
collisions.$^{\ref{cen}}$
Experimentally the situation is unclear: some, but not all, cosmic
ray experiments see Centauro behavior, while collider experiments
do not see Centauro type events.  Hopefully the ``Mini Max'' experiment
at FNAL may help settle the issue.  In the interim, it behooves theorist
to consider seriously the possibility.

I introduce the quantity
${\cal R}_3$, which is the ratio of neutral to the total number of pions.
In the average, hadronic collisions conserve isospin.  This will be
true if one averages over different events (say in $pp$ collisions),
or over all regions in rapidity of $AA$ collisions.  Thus in
the average there
will be a binomial distribution in ${\cal R}_3$,
strongly peaked about the isospin symmetric value of $1/3$.

But the behavior of the average might not be representative of all events.
Suppose that in individual events single domains are produced,
in which the pion field
has a nonzero vacuum expectation value which points in a given, fixed
direction in isospin space.  If all directions in isospin space are
equally likely, while the average value of
${\cal R}_3$ in that domain
is $1/3$, the distribution is far from binomial,
$= 1/(2 \sqrt{{\cal R}_3})$.$^{\ref{bk}}$
At this point I haven't spoken of how, in detail, such domains could
be produced dynamically; a possible mechanism is the subject of the
following.

Note that
while I have chosen the $3$ direction in isospin
to be that for neutral pions,
there is nothing special in this choice.  For
example, if ${\cal R}_1$ is the fraction of pions in the isospin
$1$ direction, then
the distribution in ${\cal R}_1$ is $1/(2 \sqrt{{\cal R}_1})$.
This is obvious, since by an isospin rotation the $3$ direction
can be relabeled as the $1$ direction.  The detailed form of the distribution
in ${\cal R}_3$ or ${\cal R}_1$ is a consequence of the symmetry group
of isospin being $O(3)$, and differs for other symmetry groups.
The important point is that the distribution
is far from binomial.

Bjorken, Kowalski, and Taylor have proposed a ``Baked Alaska'' scenario
wherein such a distribution is produced in
$pp$ collisions; the idea here is to trigger specifically
on $pp$ collisions in which the multiplicity
is very large.$^{\ref{bkt},\ref{kt},\ref{misc}}$
They refer to a domain with a fixed
direction in isospin space as
a ``Disoriented Chiral Condensate'' (DCC); it is disoriented because
in the true vacuum the pion field doesn't point in this direction (instead
the sigma field acquires a vacuum expectation value).
In some abuse of terminology, I refer to any
system which generates an isospin distribution $= 1/(2 \sqrt{{\cal R}_3})$
as that of a DCC.  It is worth remembering, however, that the way in
which DCC's can arise in hadronic collisions can, and in general will be,
very different from that in $AA$ collisions.

The behavior of central $AA$ collisions at high energies is manifestly
an example of a system with a large number of pions; for example,
at RHIC and LHC, there will be on the order of thousands of pions per
unit rapidity.  The hope is then that while the total isospin in
conserved for the total collision, that for given slices in rapidity,
the distribution in ${\cal R}_3$ is not binomial, but that of a DCC.
As emphasized
by Blaizot and Krzywicki,$^{\ref{bk}}$ the difficulty is understanding
why domains should be large in the transverse dimension: many transverse
domains, oriented in different directions, washes out the DCC effect to
give the usual binomial distribution.

Rajagopal and Wilczek$^{\ref{raw}}$ (R\&W)
proposed a dynamical mechanism for
obtaining DCC distributions in heavy ion collisions.
They use a linear sigma model to describe the chiral behavior over large
distances, and make a dramatic assumption about the dynamical evolution
in time.  They assume that the dynamics is that of a ``quench'': the
initial state is chirally symmetric, as appropriate
at high temperature, but its evolution forward in time is by means of
of the (classical) equations of motions at {\it zero} temperature.
R\&W assert that this dynamics, which
is very far from equilibrium, produces amplification and coherent oscillation
of pions at {\it low} momentum; that is, large DCC domains.

In this talk I review two attempts by S. Gavin, A. Gocksch and I
to understand how disoriented chiral condensates might arise in heavy
ion collisions.$^{\ref{ggp1},\ref{rdp2},\ref{ggp3}}$
Our work was directly motivated by the work of Bjorken, Kowalski, and
Taylor, and also by the work of Rajagopal
and Wilczek.  At the outset I must confess that while we all speak
of DCC's, in detail the mechansims of Bjorken and of Rajagopal and Wilczek
are really
very different.  In the first section we consider the work of Rajagopal
and Wilczek in detail, extending their results.$^{\ref{ggp1}}$
Unfortunately, our
results are negative: we find that the only way to get large DCC's is if
there is at least one light field about.  In the third section I
turn to what seems a long digression
on a disconnected topic: the phase diagram of $QCD$
with $2+1$ flavors.  This leads, however, to the speculation that the
equilibrium phase diagram for $2+1$ flavors could itself generate a
large distance scale.  This then {\it might} generate large domains of DCC's.

\subsection{DCC's in a two flavor model}

In this subsection I review results$^{\ref{ggp1}}$
which extend previous work
by Rajagopal and Wilczek, R\&W.$^{\ref{raw}}$
Our major purpose was to concentrate on the question of domain size and on
the experimentally relevant question of the nature of the
distribution in ${\cal R}_3$.
While our results, like those of R\&W, are the product of
numerical simulations,
much of our understanding follows from the analysis
of Boyanovsky {\it et al.};$^{\ref{boy}}$ who
consider the question of domain growth following a quench.  While the
work of (\ref{boy}) was done primarily in weak coupling (as applies
in inflationary cosmology), it is direct to extend it to strong coupling,
at least qualitatively.

For the purposes of discussion we take the
assumptions of R\&W for granted.
This means that we consider only the effects of two quark
flavors by means of a linear sigma model, ignoring vector mesons.
This involves an $O(4)$ vector $\Phi = (\sigma, \vec \pi)$ where $\sigma$ is
an isosinglet $J^P = 0^+$ field.  The action for the $\Phi$ field is
\begin{equation}
{\cal L} \; = \;
\frac{1}{2} \; (\partial_\mu \Phi)^2 \;
- \; H\cdot \Phi \; + \; \frac{\lambda}{2} v^2 \Phi^2 \; + \;
\frac{\lambda}{4} (\Phi^2)^2
{}.
\end{equation}
\label{eq:twofl}
At zero temperature
the vacuum state is $\langle \Phi \rangle = (v,\vec{0})$.
The parameters of the model are the quartic coupling
$\lambda$, the vacuum expectation value of the sigma field, $v$,
and a background magnetic field, $H = h (1,\vec{0})$,
which makes the pions massive.
To treat the theory in strong coupling we discretrize it by
putting it on a spatial lattice with a lattice spacing of
$a = 1$ fermi ($fm$).
R\&W$^{\ref{raw}}$
studied the physically relevant case of strong coupling:
$\lambda = 20$, $v = 87.4 MeV$, and $h = (119 MeV)^3$.  With
these values the pion decay constant
$f_{\pi} = 92.5 MeV $, the mass of the pion is
$m_{\pi} = 135 MeV$, while that of the sigma is
$m_{\sigma} = 600 MeV$.

(The value of $m_{\sigma} = 600 MeV$ is traditional is the two flavor
model.  For the realistic case of three flavors in the next subsection,
we shall instead identify the sigma field with a heavier particle,
near $1 GeV$ in mass.  For now I just
note that a heavier sigma just makes the coupling
even stronger than $\lambda = 20$.)

To understand the dynamics better, we also considered the model
at extremely weak coupling, $\lambda = 10^{-4}$.
In working at weak coupling we tried to keep at least some of
the physics constant by holding
the lattice spacing $a$ and the pion decay constant
$f_{\pi}$ fixed.
This implies that the pion and sigma fields become light in weak coupling:
for $\lambda = 10^{-4}$ and $f_{\pi} = 92.5 MeV$, we obtain
$v = 87.4 MeV$, $h = (2.03 MeV)^3$,
$m_{\pi} = .3 MeV$ and $m_{\sigma} = 1.8 MeV$.  In other words, weak
coupling means that the potential is flat.

We assume that the evolution of the system is ``quenched''.  The
idea is that the system cools so rapidly that the system finds
itself in a state typical of high temperature, even though the
evolution forward in time is by means of the equations of motion
at {\it zero} temperature.  We do not address the question of whether
or not this is realistic for heavy ion collisions; it certainly is
the extreme limit of a range of possibilities.
We begin with a high temperature state which is chirally symmetric,
$\langle \Phi\rangle = 0$;
to simulate the effect of fluctuations
in the high temperature initial state, we distribute the fields
as gaussian random variables with $\langle \Phi
\rangle = 0$, $\langle \Phi^2 \rangle = v^2/4$ and $\langle \dot \Phi^2
\rangle =v^2/1\,fm^2$ following R\&W.$^{\ref{raw}}$
Pion domains can form in a quench because the chirally symmetric
initial state, $\langle \Phi \rangle = (0,\vec{0})$, is unstable
against small fluctuations in the $T=0$ potential.$^{\ref{raw},\ref{boy}}$
In essence, the system ``rolls down'' from the unstable local maximum of
$V(\Phi)$ towards the nearly stable values $\Phi^2 = {f_\pi}^2$.  This
process is known in condensed matter physics as spinoidal
decomposition.$^{\ref{boy}}$  Long-lived DCC field configurations with
$\vec \pi \neq 0$ can develop during the roll-down period.
The field will eventually settle into stable
oscillations about the unique $h\neq0$ vacuum, $\langle \Phi \rangle =
(f_\pi,\vec{0})$.  Oscillations continue until interactions
eventually damp the motion through pion radiation.

In the the Hartree
approximation the equations of motion for the
Fourier components of the pion field are:$^{\ref{raw},\ref{boy}}$
\begin{equation}
{d^2 \over dt^2}{{\vec\pi}_{\vec {k}}} \; = \;
\left( \lambda (v^2 - \langle\Phi^2\rangle) - k^2 \right)
{{\vec\pi}_{\vec {k}}}\; .
\end{equation}
\label{eq:motion}
Even in the Hartree approximation we see that
field configurations with $\langle\Phi\rangle=0$,
$\langle\Phi^2\rangle < v^2$ and momentum $k < \sqrt{\lambda}v $ are
unstable, and grow exponentially.  Conversely, modes with larger momentum,
$k < \sqrt{\lambda}v $ are stable, and do not grow.
Of all of the unstable modes, the constant mode, with $k=0$, grows
the fastest; its growth has a natural time scale, which is
\begin{equation}
\tau_{\rm sp} = \{\lambda (v^2 - \langle\Phi^2\rangle)\}^{-1/2}
\sim \sqrt{2}/m_\sigma.
\end{equation}
\label{eq:time}
In strong coupling, $\lambda = 20$, this time is $.5 \, fm/c$.
The exponential growth of the unstable modes continues
until $\langle\Phi^2\rangle$ reaches $v^2$, when it begins to
oscillate about the stable vacuum.
Rajagopal and Wilczek found that the power $\propto
{\pi^a}_{ \vec {k}} {\pi^a}_{ - \vec{k}}$ in the low
momentum pion modes indeed grows when the exact classical equations of
motion are integrated for $\lambda = 20$.
Following Boyanovsky et al$^{\ref{boy}}$, one can estimate how
long it takes domains to grow; not suprisingly, this time scale
is on the order of the spinoidal times, $\tau_{sp}$.  After this
time, the fields sense that there is a stable minimum to the potential,
and so from the nonlinearities in the potential, the growth of domains
shuts off.  Consequently, if the coupling were weak,
so the sigma meson is light, then it takes
long time to roll down to the bottom of the potential, and domains
have plenty of time to grow.  For the realistic case of strong coupling,
however, the rolldown is very rapid, and so the domains are small,
$\sim 1 fm$.  This is what one would expect from a system in which
the Compton wavelength of the pion at rest is of order $\sim 1 fm$.

We confirmed this intuition through numerical simulations.
An asymmetric lattice geometry of dimensions
$10^2 \times 40$ was chosen
in order to study the issue of domain size in the one, long direction;
we introduce the average of the pion field over the transverse
dimensions,
$\vec{\pi}_L(z,t) = \sum_{x,y}\vec{\pi}(x,y,z,t)/(10)^2$.
Initially, $\pi_L$ starts out completely random.  In weak coupling,
by times of $100 fm$ domains ---
regions in which the field is slowly varying about
some nonzero value --- are evident.  In contrast,
in strong coupling by times of $100 fm$ the pion field
is always quite random, oscillating with small amplitude about zero.
We found that the pion correlation function is long
ranged in weak coupling, but short ranged
in strong coupling.$^{\ref{rdp2},\ref{ggp3}}$

The most important quantity is the distribution in ${\cal R}_3$.
We assume that we can extract this ratio from
the classical pion fields in the most naive way possible, by computing
${\cal R}_3=\langle(\pi^3)^2\rangle/
(\langle(\pi^1)^2\rangle + \langle(\pi^2)^2\rangle
+ \langle(\pi^3)^2\rangle)$.
Histograms of ${\cal R}_3$ were obtained by evolving $200$ independent
configurations forward in time to $t = 150$ fm in weak coupling,
and times of $t=30$ fm in strong coupling.
In weak coupling the distribution is far from binomial, and is
approaching the DCC value of $1/(2 \sqrt{{\cal R}_3})$.
In strong coupling, however, the distribution is clearly binomial,
peaked about the expected value of $1/3$.

We have performed numerical simulations at other couplings to see
where the crossover from a binomial to a DCC distribution
occurs.$^{\ref{rdp2}}$
For our lattice of $10^2 \times 40$,
it appears that a DCC distribution only arises at
very weak coupling, $\lambda \sim 10^{-2}$, as expected from
the domain size estimate of (2).

In summary, in a realistic two flavor model we find that following
a quench, DCC's are only produced in small, pion sized domains.
There appears to be one way for large DCC's to be produced, and that
is if there is a light particle about, whose Compton wavelength automatically
provides a size for the DCC.  But a quench alone does not seem to give
us a large distance scale.

\subsection{Three flavors}

In the next subsections I analyze
the phase diagram of $QCD$.$^{\ref{rdp2}}$
This is similar to the analysis of sec. 2, except that I
work not with $2$ but the realistic case of $2+1$ flavors.  The price
I pay is that the contribution of the vector fields is dropped.
Again, implicitly I speak of nonzero temperature; what happens at
nonzero baryon density (if the temperature is small)
may well be very different.

{}From the previous discussion, remember that for three colors,
the order of the phase transition appears to depend crucially on the
values of the quark masses.  This is definitely special to three colors.
For example, consdier the
limit of infinitely many colors, $N_c \rightarrow
\infty$.  The basic physics can be understood simply by remembering
that there are of order $\sim N^2_c$ gluons
versus order $\sim N_c$ quarks.
Now assume that the deconfining transition is of first order at
$N_c = \infty$; now while I certainly cannot prove it, the most natural
possibility is that the first order deconfining transition always
dominates the chiral transition, so that both take place at the
same temperature.$^{\ref{largen}}$  This need not be the case, for
it is logically possible for the chiral transition to occur at a higher
temperature than the deconfining transition.  (The opposite possibility,
of a deconfining temperature which is higher than the chiral transition,
probably contradicts general theorems on realizations of chiral symmetry,
but as of yet no strict proof has been given.)  Nevertheless, as the
deconfining transition the free energy goes from being of order one,
from hadrons, to of order $N_c^2$, from gluons, and it is difficult to
imagine that this in and of itself doesn't trigger the chiral transition
at the same time.

Thus we see that at least in this instance, three colors is {\it not}
close infinity:$^{\ref{karsch},\ref{col}}$
for three colors and ``$2+1$'' flavors, while the transition appears
to be of first order for the deconfining transition without dynamical
quarks, ${\cal M} = \infty$ and for the chiral limit, ${\cal M} = 0$,
these lines do not meet --- there is a gap, with $QCD$ somewhere
in between.

Lines of first order transitions typically end in critical points, so it
is natural to ask about the two critical points.
Consider first working down from infinite quark mass.
As ${\cal M}$ decreases from
${\cal M}=\infty$, the line of deconfining first order
transitions ends in a deconfining critical point.
By an analysis similar to that
in sec. (3.2) below, one can show that at the deconfining critical
point, correlation functions between Polyakov
lines are infinite ranged, and that it
lies in the universality class of the Ising model, or a $Z_2$
spin system, in three dimensions.

The other limit is to go up from zero quark mass.  For three flavors
the chiral phase transition is of first order at $m=0$, so as $m$ increases,
the line of first order transitions can end in a chiral critical point.
In the following we show that for $2+1$ flavors,
at the chiral critical point the only massless field is the sigma
meson.  Thus again the chiral critical point lies in
the universality class of the Ising model, or a $Z(2)$ spin system.

We stress, however, very different fields becomes massless at each
critical point; the two critical points are not naturally related to
each other.  At least for three colors this seems to be the case, as
there is a large gap in between the two critical points, in which there
is no true phase transition, only a smooth crossover.  Nevertheless,
our results indicate that even in the region of smooth crossover ---
which includes $QCD$ --- that very interesting and unexpected physics
can be going on.

Because we are interested in making contact with $QCD$, we begin
in the next subsection by
fitting the scalar and pseudoscalar mass spectrum in $QCD$
to the results found in a linear sigma model.
This is an old story$^{\ref{chan}}$ and relatively straightforward.
The classification of the chiral critical point then follows directly.
When we then try to connect the two, however, we find a surprise.
Numerical simulations of lattice gauge theory$^{\ref{col}}$
indicate that as a function of the mass parameter ${\cal M}$,
$QCD$ is not far from the chiral critical point.
We show that if $QCD$ is in fact near the chiral critical point,
then that, and the nature of the fit to the zero temperature spectrum,
makes very interesting predictions about the behavior of $QCD$
at nonzero temperature.

\subsubsection{The QCD spectrum at zero temperature}

In this subsection we use a linear sigma model to model the full theory
with three quark
flavors.$^{\ref{pw},\ref{gold},\ref{chan},\ref{mo1},\ref{mo2}}$
We ignore the contribution of vector mesons, which will be incorporated
at a later date.$^{\ref{rdp3}}$
This type of analysis was first
carried out by Chan and Haymaker;$^{\ref{chan}}$
recent analyses, with similar results, are given by Meyer-Ortmanns, Pirner,
and Patkos,$^{\ref{mo1}}$ and by
Metzger, Meyer-Ortmanns, and Pirner.$^{\ref{mo2}}$
For three quark flavors the sigma field $\Phi$ is a complex valued,
three by three matrix, proportional to the quark fields
as $\Phi \sim \overline{q}_{left} \; q_{right}$.  In flavor space it
can be decomposed as
\begin{equation}
\Phi \; = \sum^{8}_{a=0} \;
\left( \sigma_a \; + \; i \, \pi_a \right) t^a  \; .
\label{eb1}
\end{equation}
($t_{1 \ldots 8}$ are the generators of the $SU(3)$ algebra in
the fundamental representation, while $t_0$ is proportional to the
unit matrix.  Normalizing the generators as $tr(t_a t_b) = \delta^{a b}/2$,
$t_0 = {\bf 1}/\sqrt{6}$.)
The fields $\sigma_a$ are components of a
scalar ($J^P = 0^+$) nonet, those of $\pi_a$ a pseudoscalar
($J^P = 0^-$) nonet.  The latter are familiar:
$\pi_{1,2,3}$ are the three pions, denoted as $\pi$ without
subscript; the
$\pi_{4,5,6,7}$ are the four kaons, the $K$'s,
while $\pi_8$ and $\pi_0$ mix to form the
mass eigenstates of the $\eta$ and $\eta'$ mesons.
Following our notations for two flavors,
we define the components of the scalar
nonet analogously: we refer to $\sigma_{1,2,3}$ as the $\sigma_\pi$'s,
to $\sigma_{4,5,6,7}$ as the $\sigma_K$'s, while $\sigma_8$ and $\sigma_0$
mix to form the $\sigma_\eta$ and $\sigma_{\eta'}$.
What is the correct
identification of the $\sigma$ meson for three flavors will be one of
the principal questions in the following.

This multiplicity of eighteen fields is to be contrasted with the
usual two flavor sigma model considered in the previous subsection,
which had only three $\pi$'s and one $\sigma$ meson.  The increase
in the number of fields
is because we are allowing for $U_a(1)$ rotations; to describe two
flavors in this way requires eight fields: the three $\pi$'s, $\sigma_\eta$,
the three $\sigma_\pi$'s, and the $\eta$.  For $n_f$ flavors, then,
in general $2 n_f^2$ fields are required to describe the lowest lying
scalar and pseudoscalar multiplets.

The effective lagrangian for the chiral $\Phi$ field is taken to be
\begin{equation}
{\cal L} \; = \; tr \left| \partial_\mu \Phi \right|^2
\; - \; tr \left( H (\Phi + \Phi^\dagger) \right)
\; + \;  \mu^2 \; tr\left(\Phi^\dagger \Phi\right)
\label{eb2}
\end{equation}
$$
\; - \; \sqrt{6} \; c \left( det(\Phi) \, + \, det(\Phi^\dagger) \right)
\; + \; (g_1 - g_2) \left( tr \Phi^\dagger \Phi \right)^2
\; + \; 3 \, g_2 \; tr\left(\Phi^\dagger \Phi\right)^2 \;
$$
This is very much like the lagrangian for two flavors in (1), except
that now new couplings are possible.
This lagrangian involves six parameters: two for
the background field $H$ (one for the up equals the down quark mass, one
for the strange quark mass), the mass parameter $\mu^2$, an ``instanton''
coupling constant $c$, and two quartic couplings, $g_1$ and $g_2$.
At large $\Phi$ the potential is bounded from below if the quartic
couplings satisfy $g_1 \geq 0$ and $g_1 + 2 g_2 \geq 0$.
(The bounds on the quartic couplings can be understood by taking
two extreme limits for $\Phi$:
the first is $\Phi$ proportional to the unit matrix, the second
is where $\Phi$ only has one diagonal element which is nonzero.)

To understand the symmetries of this lagrangian, consider
the theory becomes less symetric as various couplings increase from zero.
When $H = c = g_2 = 0$,
the theory only involves the quadratic invariant
$tr(\Phi^\dagger \Phi) = \sum^8_{a=0} (\sigma_a^2 + \pi_a^2)$,
and so has the very large symmetry group of $O(18)$.
When $g_2 \neq 0$, the symmetry reduces
to $SU(3) \times SU(3) \times U(1)$, with the $U(1)$ that for
axial fermion number.  The effects of instantons, or more generally
topological fluctuations and the
Adler-Bell-Jackiw anomaly, generate a nonzero value for the instanton
coupling constant $c$, and reduces the symmetry to $SU(3) \times
SU(3)$.  Lastly, spontaneous symmetry breaking,
where $\langle \Phi \rangle \sim {\bf 1}$, occurs if the mass squared,
$\mu^2$ is negative; actually, due to the determinental term, spontaneous
symmetry breaking can even occur for some values of positive $\mu^2$.
Since $\langle \Phi \rangle \sim {\bf 1}$, the symmetry is then
reduced to the usual $SU(3)$ symmetry of isospin plus strangness.

For the background field $H$ we take
$H = h_0 \, t_0 - \sqrt{2} \, h_8 \, t_8$.
Working out the algebra, $h_0$ and $h_8$ are proportional to the
current quark masses as
\begin{equation}
{\cal M}_{up} \; \equiv \; {\cal M}_{down} \; \sim \; h_0 - h_8
\;\; , \;\; {\cal M}_{strange} \sim h_0 + 2 \, h_8 \; .
\label{eb3}
\end{equation}
With the octet magnetic field $h_8 \neq 0$, then,
only remaining symmetry is merely the usual
$SU(2)$ symmetry of isospin.  Effects
from isospin breaking, as from ${\cal M}_{up}
\neq {\cal M}_{down}$, are negligible
for our purposes, since every other mass scale is so much larger.
(Not to mention the inclusion of electromagnetic effects.)
Note that $H$ has dimensions of mass cubed; hence two powers
of some fixed $QCD$ scale enter to make up the mass
dimensions in (\ref{eb3}).  I assume that these two powers of a
$QCD$ scale are truly fixed, and do {\it not} vary with, say, temperature.
This appears to give a behavior with nonzero temperature which is consistent
with other analyses, but I must confess that I do not not know how to
derive the value of the background field from the fundamental theory in
a way in which these two powers of the $QCD$ scale would be manifest.n

The analysis of the effective lagrangian is uniformly at the simplest level
of mean field theory.
I assume that in the true vacuum
there are nonzero expectation values for $\sigma_8$
and $\sigma_0$,
\begin{equation}
\sigma_0 \; \rightarrow \; \Sigma_0 \, + \, \sigma_0 \;\; , \;\;
\sigma_8 \; \rightarrow \;
- \, \sqrt{2} \; \Sigma_8 \, + \, \sigma_8 \; ,
\label{eb4}
\end{equation}
and then expand in powers of $\sigma_a$ and $\pi_a$.  To linear order
there are two equations of motions, which fixes the values
the two vacuum expectation values, $\Sigma_0$ and $\Sigma_8$, as
$$
h_0 \; = \; \, M^2 \, \Sigma_0
\, - \, c \, \left( \Sigma_0^2 - \Sigma^2_8 \right)
\, + \, 4 g_2 \, \Sigma^2_8
\left(\Sigma_0 + \frac{1}{2} \, \Sigma_8 \right)
\; ,
$$
\begin{equation}
h_8 \; = \; \, M^2 \, \Sigma_8
\, + \, c \, \Sigma_8 \, \left( \Sigma_0 - \Sigma_8 \right)
\, + \, 2 g_2 \, \Sigma_8 \, \left(\Sigma_0 + \Sigma_8\right)
\, \left(\Sigma_0 + \frac{1}{2} \, \Sigma_8 \right) \; .
\label{eb5}
\end{equation}
A second mass parameter, $M^2$, has been introduced; it is related to
the $\mu^2$ of (\ref{eb2}) as
\begin{equation}
M^2 \; = \; \mu^2
\, + \, g_1 \, \left(\Sigma^2_0 \, + \, 2 \, \Sigma^2_8\right) \; .
\label{eb6}
\end{equation}
Expansion of the effective lagrangian to quadratic order gives the masses
of the pseudoscalar octet,
$$
m^2_\pi \; = \; M^2 \, - \, c \, \left( \Sigma_0 + 2 \Sigma_8 \right)
\, - \, 2 g_2 \, \Sigma_8 \,
\left(\Sigma_0 + \frac{1}{2} \, \Sigma_8 \right)
\; = \; \frac{h_0 - h_8}{\Sigma_0 - \Sigma_8} \; ,
$$
$$
m^2_K \; = \; M^2 \, - \, c \, \left(\Sigma_0 - \Sigma_8 \right)
\, + \, g_2 \, \Sigma_8 \left(\Sigma_0 + 5 \Sigma_8\right)
\; = \; \frac{h_0 + h_8/2}{\Sigma_0 + \Sigma_8/2} \; ,
$$
$$
m^2_{\pi_8} \; = \; M^2 \, - \, c \left(\Sigma_0 - 2 \Sigma_8\right)
\, + \, 2 g_2 \, \Sigma_8 \,
\left(\Sigma_0 + \frac{1}{2} \, \Sigma_8 \right) \; ,
$$
$$
m^2_{\pi_8 \pi_0} \; = \; \sqrt{2} \, \Sigma_8
\, \left( c \, - \, 2 g_2
\left( \Sigma_0 + \frac{1}{2} \, \Sigma_8 \right) \right) \; ,
$$
\begin{equation}
m^2_{\pi_0} \; = \; M^2 \, + \, 2 c \, \Sigma_0 \; .
\label{eb7}
\end{equation}
Notice that $m^2_{\pi_8}$, $m^2_{\pi_8 \pi_0}$, and $m^2_{\pi_0}$ refer
to the octet and singlet parts of the pseudoscalar nonet, which mix to
give the masses of the $\eta$ and $\eta'$,
$m_\eta$ and $m_{\eta'}$.  In the chiral limit, $h_0 = h_8 = 0$,
the chiral condensate is $SU(3)$ symmetric, $\Sigma_0 \neq 0$ and
$\Sigma_8 = 0$.  In this limit it is easy to see that the $\pi$, $K$, and
$\eta$ mesons are all massless; the $\eta'$ meson remains massive
because of the instanton coupling $c$.

For the scalar nonet,
$$
m^2_{\sigma_\pi} \; = \; M^2
\, + \, c \, \left( \Sigma_0 + 2 \Sigma_8 \right)
\, + \, 2 g_2 \, \left(\Sigma_0^2 - 3 \Sigma_0 \Sigma_8
+ \frac{1}{2} \, \Sigma_8 \right) \; ,
$$
$$
m^2_{\sigma_K} \; = \; M^2
\, + \, c \, \left(\Sigma_0 - \Sigma_8\right)
\, + \, 2 g_2 \, \left(\Sigma_0 + \Sigma_8 \right)
\left(\Sigma_0 + \frac{1}{2} \, \Sigma_8\right) \; ,
$$
$$
m^2_{\sigma_8} \; = \; \mu^2
\, - \, c \, \left(\Sigma_0 - 2 \Sigma_8\right)
\, + \, g_1 \, \left( \Sigma_0^2 + 6 \Sigma_8^2 \right)
\, + \, 2 g_2 \, \left(\Sigma_0^2 + 3 \Sigma_0 \Sigma_8
+ \frac{3}{2} \, \Sigma_8^2 \right) \; ,
$$
$$
m^2_{\sigma_8 \sigma_0} \; = \; - \, \sqrt{2} \, \Sigma_8
\, \left( c \, + 2 g_1 \, \Sigma_0
\, + \, g_2 \left( 4 \Sigma_0 + 3 \Sigma_8 \right) \right) \; ,
$$
\begin{equation}
m^2_{\sigma_0} \; = \;  \mu^2 \, - \, 2 \, c \, \Sigma_0
\, + \, g_1 \, \left( 3 \Sigma_0^2 + 2 \Sigma_8^2 \right)
\, + \, 4 g_2 \, \Sigma_8^2 \; .
\label{eb8}
\end{equation}
For the scalars, $m^2_{\sigma_8}$, $m^2_{\sigma_8 \sigma_0}$, and
$m^2_{\sigma_0}$ mix to give $m_{\sigma_{\eta}}$ and $m_{\sigma_{\eta'}}$.
All particles in the scalar nonet are expected to remain massive even in
the chiral limit.

We also need the pion and kaon decay constants,
\begin{equation}
f_\pi \; = \; \sqrt{\frac{2}{3}} \; \left( \Sigma_0 - \Sigma_8 \right)
\;\; , \;\;
f_K \; = \; \sqrt{\frac{2}{3}} \;
\left( \Sigma_0 + \frac{1}{2} \, \Sigma_8 \right) \; .
\label{eb9}
\end{equation}

\begin{figure}[t]
\centerline{\epsffile{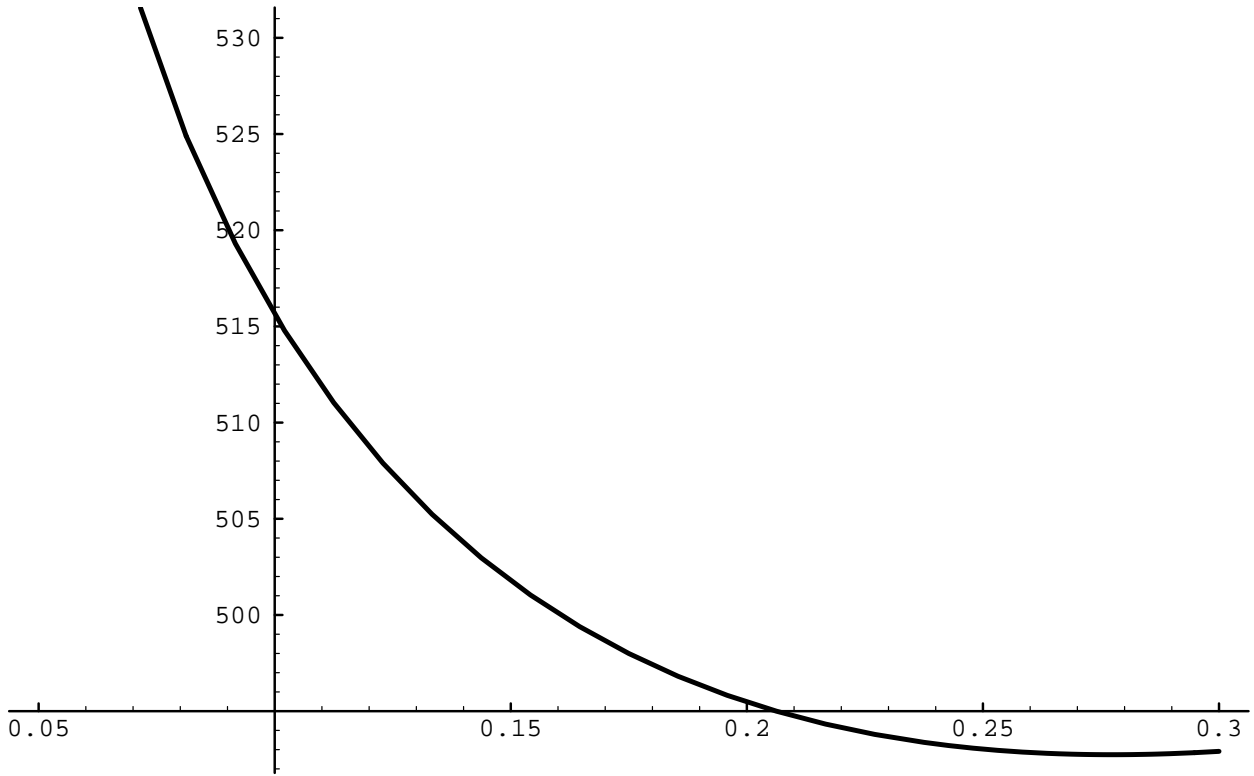}}
\caption{
The kaon mass, $m_K$, in $MeV$, versus the ratio of
octet to singlet vacuum expectation values, $s = \Sigma_8/\Sigma_0$.
The kaon mass is rather insensitive to $s$.
}
\end{figure}

The expressions in (\ref{eb7})--(\ref{eb9}) agree with previous
results,$^{\ref{chan},\ref{mo1},\ref{mo2}}$
up to differences in normalization of the coupling constants
and vacuum expectation values in (\ref{eb2}) and (\ref{eb4}).  There is one
unexpected detail:
naively one would expect that the mass $\mu$ and the quartic
coupling $g_1$ would enter independently.  In fact, for the equations of
motion in (\ref{eb5}), the masses of the entire pseudoscalar nonet
in (\ref{eb7}), and the masses of half the scalar nonet --- for the
$\sigma_{\pi}$ and the $\sigma_{K}$ in (\ref{eb8}) --- these two parameters
only enter in a certain combination, through the
single parameter $M^2$ of (\ref{eb6}).
This means that we can fit to the pseudoscalar spectrum, and so fix $M^2$,
and yet still be free to vary $\mu^2$ (or conversely $g_1$): the {\it only}
change will be to alter the masses of the $\sigma_{\eta}$ and the
$\sigma_{\eta'}$.  This technical detail will play an important role in
what follows; although
there must be some simple group theoretic reason for it,
I don't know what it is.

There is some freedom in deciding how to fit the parameters of the linear
sigma model.  Various kinds of fits are given by Meyer-Ortmanns, Pirner,
and Patkos,$^{\ref{mo1}}$ and by Metzger,
Meyer-Ortmanns, and Pirner.$^{\ref{mo2}}$
Following the experience of Chan and Haymaker$^{\ref{chan}}$ we do not
fit to the entire pseudoscalar mass spectrum for the $\pi$, $K$, $\eta$,
and $\eta'$ mesons, since
it turns out that the kaon mass is fairly insensitive to the
ratio of vacuum expectation values, $\Sigma_8/\Sigma_0$.

Because of this,
we leave the ratio $\Sigma_8/\Sigma_0$ as a free parameter, and fit just
to the pion decay constant $f_\pi$ and to
to the masses for the $\pi$, $\eta$, and $\eta'$ mesons.  I assume
the values:
\begin{equation}
f_{\pi} \; = \; 93 \, MeV \;\; , \;\;
m_\pi \; = \; 137 \, MeV \;\; , \;\;
m_\eta \; = \; 547 \, MeV \;\; , \;\;
m_{\eta'} \; = \; 958 \, MeV \; .
\label{eb10}
\end{equation}

\begin{figure}[t]
\centerline{\epsffile{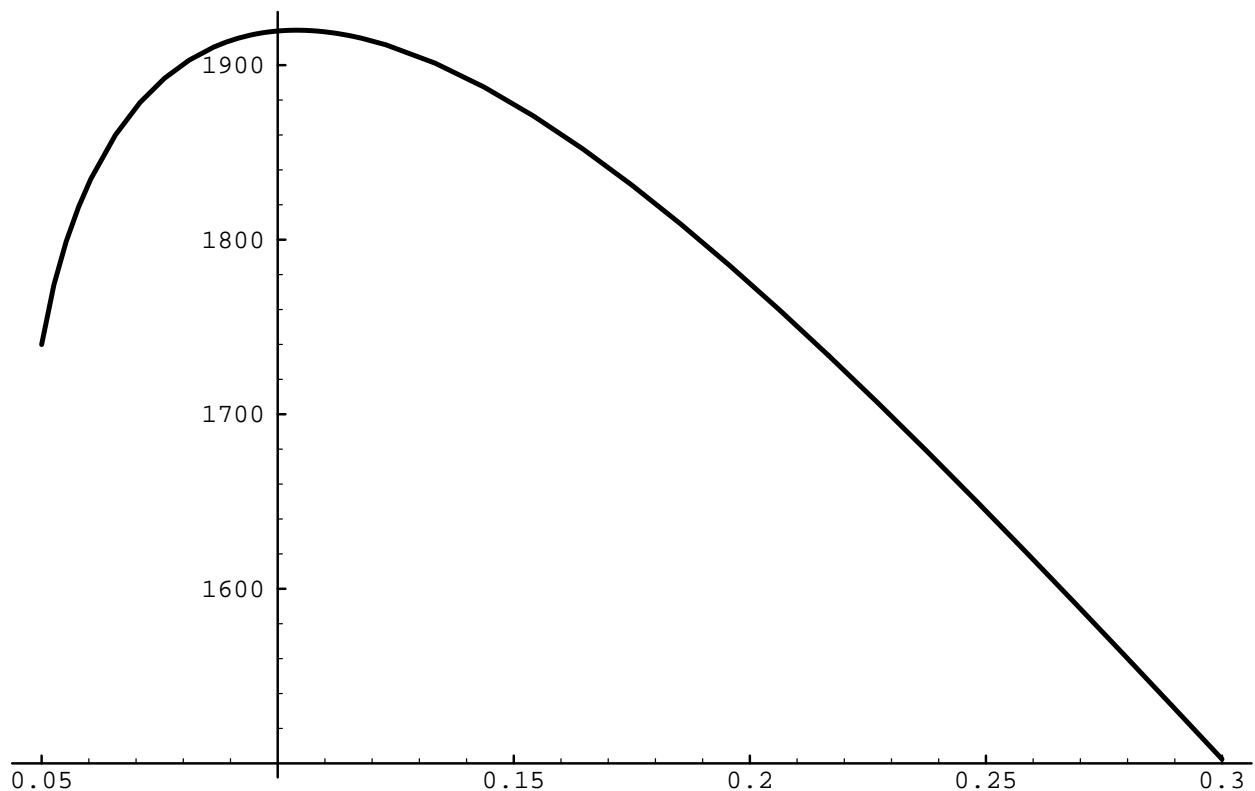}}
\caption{
Coupling for the anomaly, $c$, in $MeV$, versus the ratio of
octet to singlet vacuum expectation values, $s = \Sigma_8/\Sigma_0$.
This coupling constant is rather insensitive to $s$.
}
\end{figure}
Then one can examine the sensitivity of various quantities to
$s = \Sigma_8/\Sigma_0$.  This is easiest of understand graphically:
in fig. (6) I show how the kaon mass changes with $s$: for small $s$
it is a little high, but not terribly so.
In fig. (7) I show how
the coupling for the anomaly changes with $s$; again, there is some variation
but nothing dramatic.

In fig. (8) I show how the coupling constant $g_2$ of the linear sigma
model changes with $s$; we see that for large $s \sim .3$ the coupling
$g_2$ becomes rather small.  Analysis shows that while $g_2$ becomes
negative for $s > .3$, its value is very small, $\leq 1.$, so that for
any reasonable values of $g_1$, the potential remains bounded from below.
Thus values of $s \leq 1.$ cannot be excluded on such grounds.
\begin{figure}[t]
\centerline{\epsffile{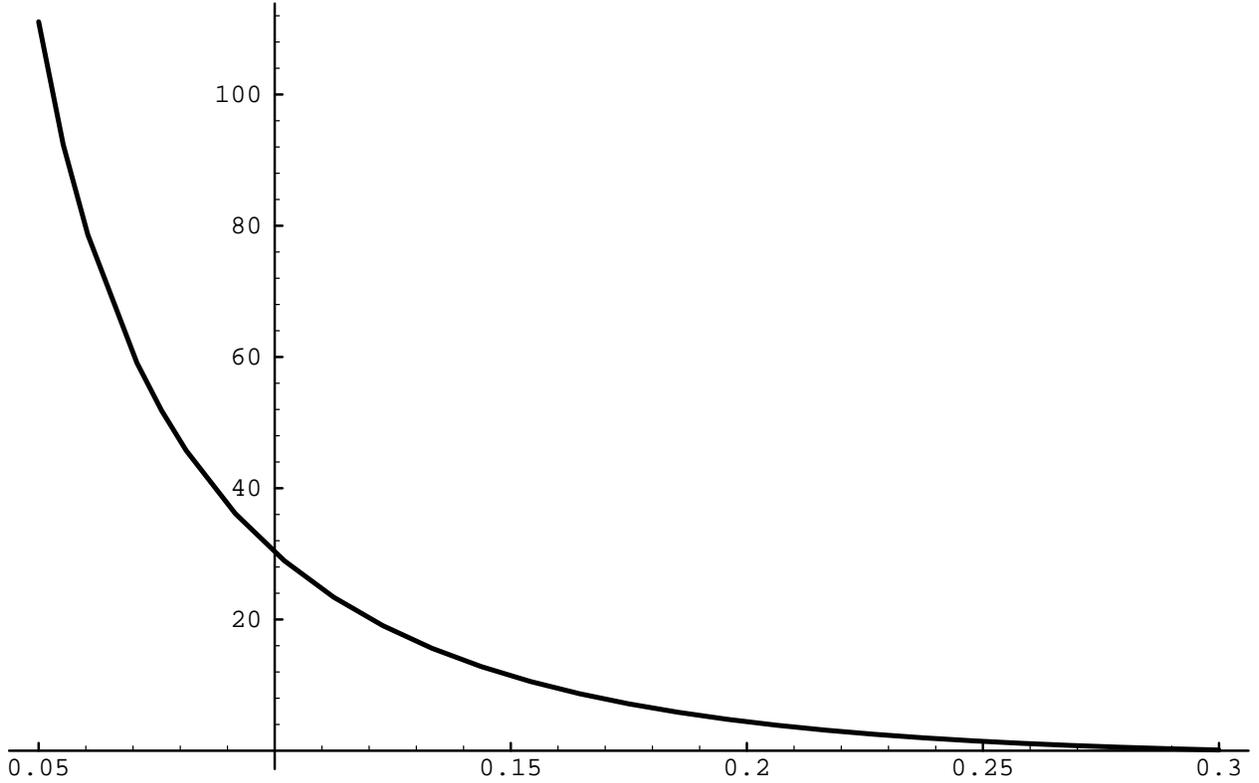}}
\caption{
The coupling $g_2$ of the linear sigma model versus the ratio of
octet to singlet vacuum expectation values, $s = \Sigma_8/\Sigma_0$.
While $g_2$ becomes much smaller for large $s$, this is rather
innocuous.
}
\end{figure}

\begin{figure}[t]
\centerline{\epsffile{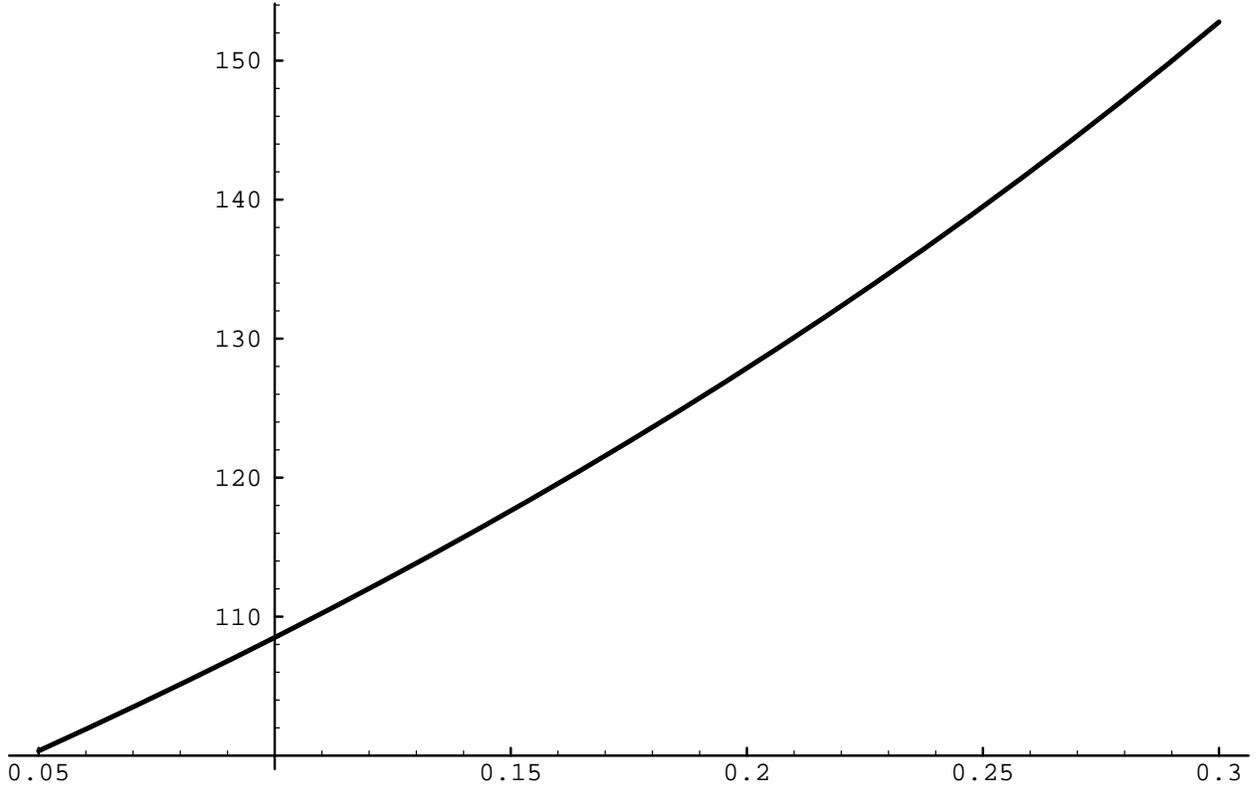}}
\caption{
Kaon structure constant, $f_K$, in $MeV$, versus the ratio of
octet to singlet vacuum expectation values, $s = \Sigma_8/\Sigma_0$.
Small values of $s$ are favored, $s \sim .1$.
}
\end{figure}

There are two quantities, however, which are {\it very} sensitive to
the value of the octet to singlet vacuum expectation values, $s$.
The first is the kaon decay constant, $f_K$.  This is illustrated in
fig. (9), where we see that small values of $s$ are preferred:
the experimental value is $f_K = 113 \, MeV$.$^{\ref{pdt}}$
Since $f_K$ isn't too far from $f_\pi$, given the definition of
$f_K$ in (\ref{eb9}), it is obvious why a small ratio of
$\Sigma_8/\Sigma_0$ is preferred.
\begin{figure}[t]
\centerline{\epsffile{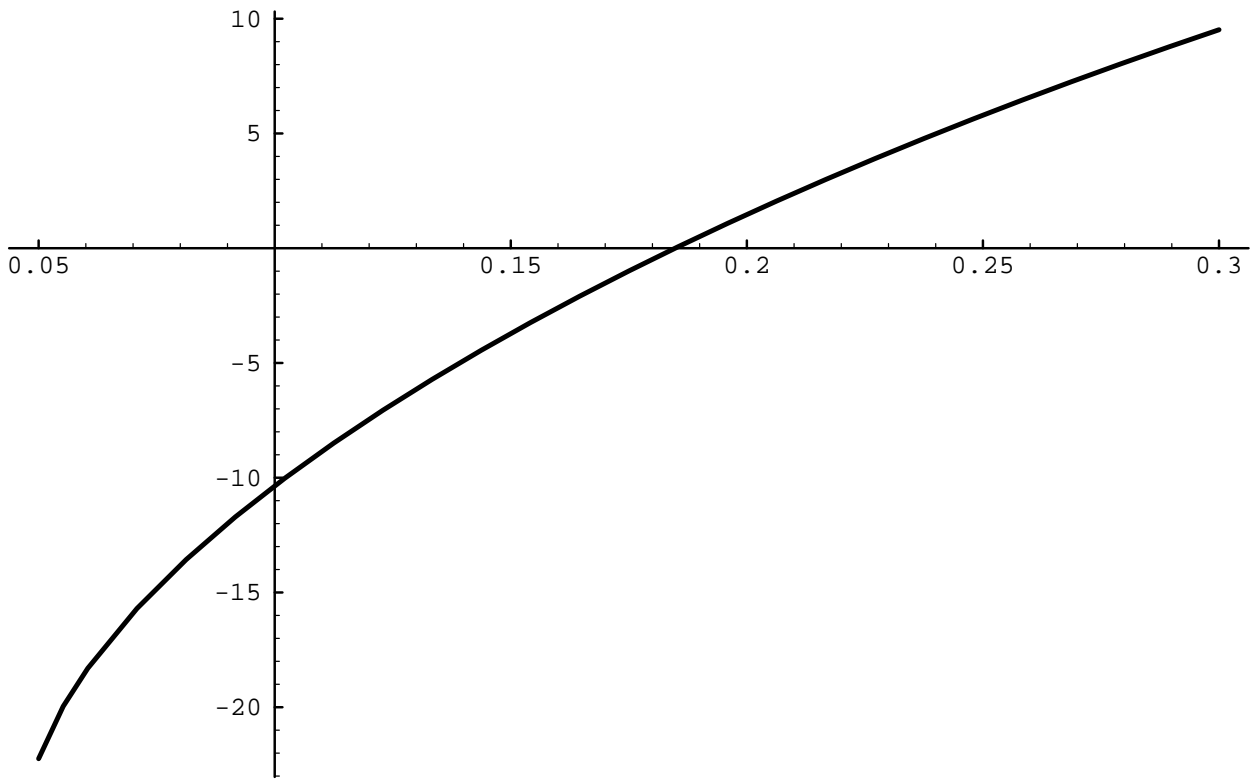}}
\caption{
The mixing angle between the $\eta$ and the $\eta'$, $\theta_{\eta \eta'}$,
versus the ratio of
octet to singlet vacuum expectation values, $s = \Sigma_8/\Sigma_0$.
Small values of $s$ are favored, $s \leq .1$.
}
\end{figure}
The second quantity is the mixing angle between the $\eta$ and the
$\eta '$, $\theta_{\eta \eta'}$.  Experimentally the situation is
a bit unclear: radiative decays favor a value of
$\theta_{\eta \eta'} = - \, 20^{o}$, while fits to the mass spectrum
favor $\theta_{\eta \eta'} = - \, 10^{o}$.  From fig. (10), however,
we see that this uncertainty doesn't really affect our fit, for either
value favors {\it low} values of $s$.
{}From fig. (9) and (10), we find that a good fit is obtained for $s = .1$.
The parameters of this solution are
$$
M^2 \; = \; + \, (642 \, MeV)^2 \;\; , \;\;
c \; = \; 1920 \, MeV \;\; , \;\;
g_2 \; = \; 30. \;\; , \;\;
$$
\begin{equation}
\Sigma_0 \; = \; 127 \, MeV \;\; , \;\;
\Sigma_8 \; = \; 13 \, MeV \;\; , \;\;
h_0 \; = \; (290 \, MeV)^3 \;\; , \;\;
h_8 \; = \; (281 \, MeV)^3 \; .
\label{eb11}
\end{equation}
For this fit with $s = \Sigma_8/\Sigma_0 \sim .1$, while
\begin{equation}
m_K \; = \; 516 \, MeV \;\; , \;\;
f_K \; = \; 109 \, MeV \;\; , \;\;
\theta_{\eta \eta'} \; = \; - \, 10.4^{o} \; .
\label{eb12}
\end{equation}
The kaon mass is a bit high, $m_K = 516 \, MeV$ instead of the (average)
experimental value of $497 \, MeV$.

For the values of (\ref{eb11}), from (\ref{eb3})
the ratio of the strange to up ($=$ down) quark masses is
\begin{equation}
\frac{{\cal M}_{strange}}{{\cal M}_{up}} \; =
\; \frac{(h_0 + 2 h_8)}{(h_0 - h_8)} \; = \; 32 \; .
\end{equation}
This is larger
than the often quoted value of $\sim 20$ because we have explicitly
allowed for a vacuum expectation value that is not $SU(3)$ symmetric,
$\Sigma_8 \neq 0$.

If one fits to the kaon mass, then $\Sigma_8/\Sigma_0$ is constrained to be
$= .19$, with $c = 1790 \, MeV$ and $g_2 = 4.98$;
the results for $f_K$ and $\theta_{\eta \eta'}$ are worse,
$f_K = 127\, MeV$, while
$\theta_{\eta \eta'} = + \, 1.^{o}$.
The freedom to play with the ratio $\Sigma_8/\Sigma_0$ and its effects
will be discussed elsewhere.$^{\ref{ggp3}}$

The values in (\ref{eb11}) give unique predictions for two masses in the
scalar nonet:
\begin{equation}
m_{\sigma_\pi} \; = \; 1177 \, MeV \;\; , \;\;
m_{\sigma_K} \; = \; 1322 \, MeV \; .
\label{eb13}
\end{equation}
There are observed states$^{\ref{pdt}}$ with these quantum numbers: the
$a_0(980)$ and the $K^*_0(1430)$, respectively.

For the purposes of illustration, in fig. (11) I illustrate how
the masses of the $\sigma_\pi$ and the $\sigma_K$ change with
$s$.  For small values of $s$, the $\sigma_K$ is heavier than the
$\sigma_\pi$, as naive intuition would suggest.  What is amusing
is that for $s \geq .2$, the $\sigma_K$, which after all is strange,
becomes lighter than the $\sigma_\pi$.  For other reasons, I do not
favor such large values of $s$, but it is interesting that such a
flip of mass values can occur.
\begin{figure}[t]
\centerline{\epsffile{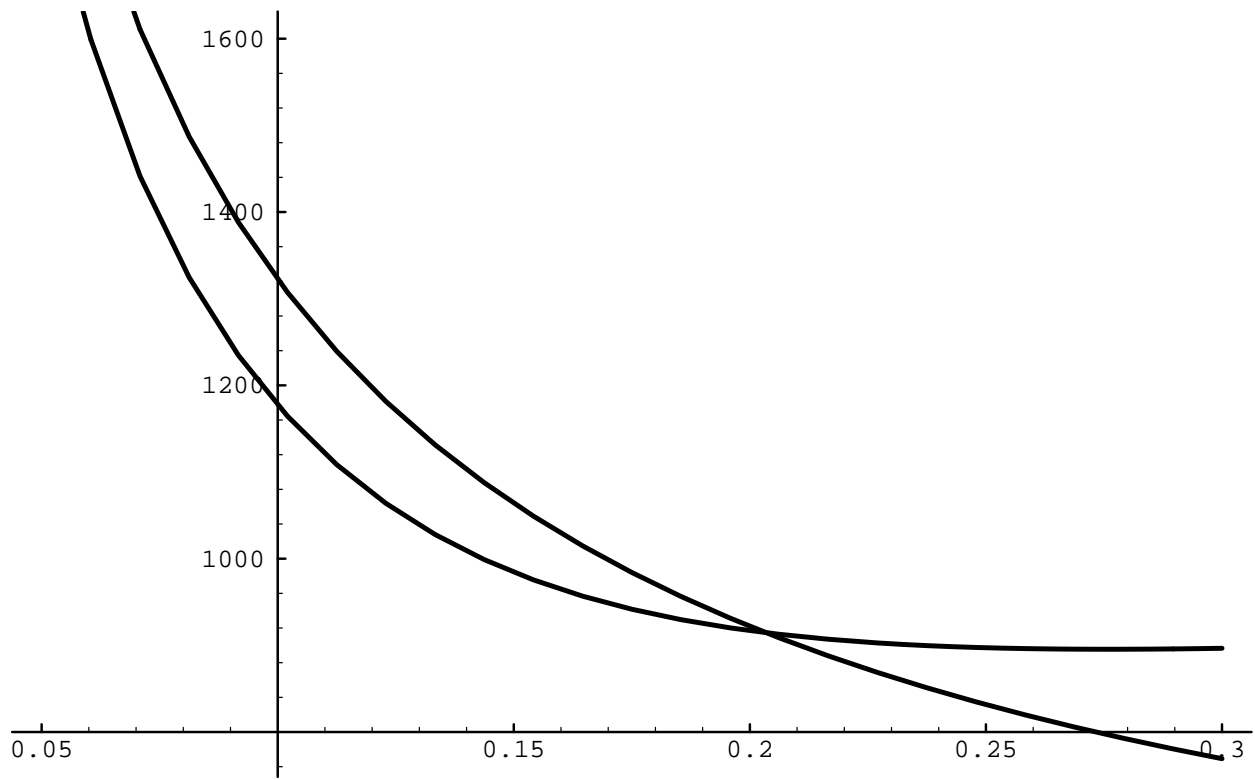}}
\caption{
The masses of the scalar pion, $\sigma_\pi$, and the scalar kaon,
$\sigma_K$, versus the ratio of
octet to singlet vacuum expectation values, $s = \Sigma_8/\Sigma_0$.
}
\end{figure}

The identification of the $a_0$ with the $\sigma_\pi$ is not
obvious (see pg. VII.21 of ref. [\ref{pdt}]).
It is reasonable to expect that the width of
the $\sigma_\pi$ should be large, while the $a_0$ is observed to have
a small width $\sim 57 MeV$.
(The $K^*_0$ is a broad resonance, with a width $\sim 287 \, MeV$.)
While this may result from a large mixing with other
channels, such as $K\overline{K},^{\ref{torn}}$ alternately it is possible
is that the $a_0 (980)$ is not what we term the
$\sigma_\pi$, but a kind of
$K \overline{K}$ molecule.$^{\ref{kkbar}}$  If true,
presumably the state which we term the $\sigma_\pi$ is
heavier than $1 \, GeV$.$^{\ref{torn},\ref{pdt}}$

There is no unique prediction for the masses of the
remaining members of the scalar
nonet, the $\sigma_{\eta}$ and the $\sigma_{\eta'}$.   As remarked,
the masses of all other fields only depends upon the
parameter $M^2$ in (\ref{eb6}).  Thus at constant $M^2$ we can vary
$g_1$ and change {\it only}
$m_{\sigma_{\eta}}$ and $m_{\sigma_{\eta'}}$.
This is illustrated in fig. (12).
\begin{figure}[t]
\centerline{\epsffile{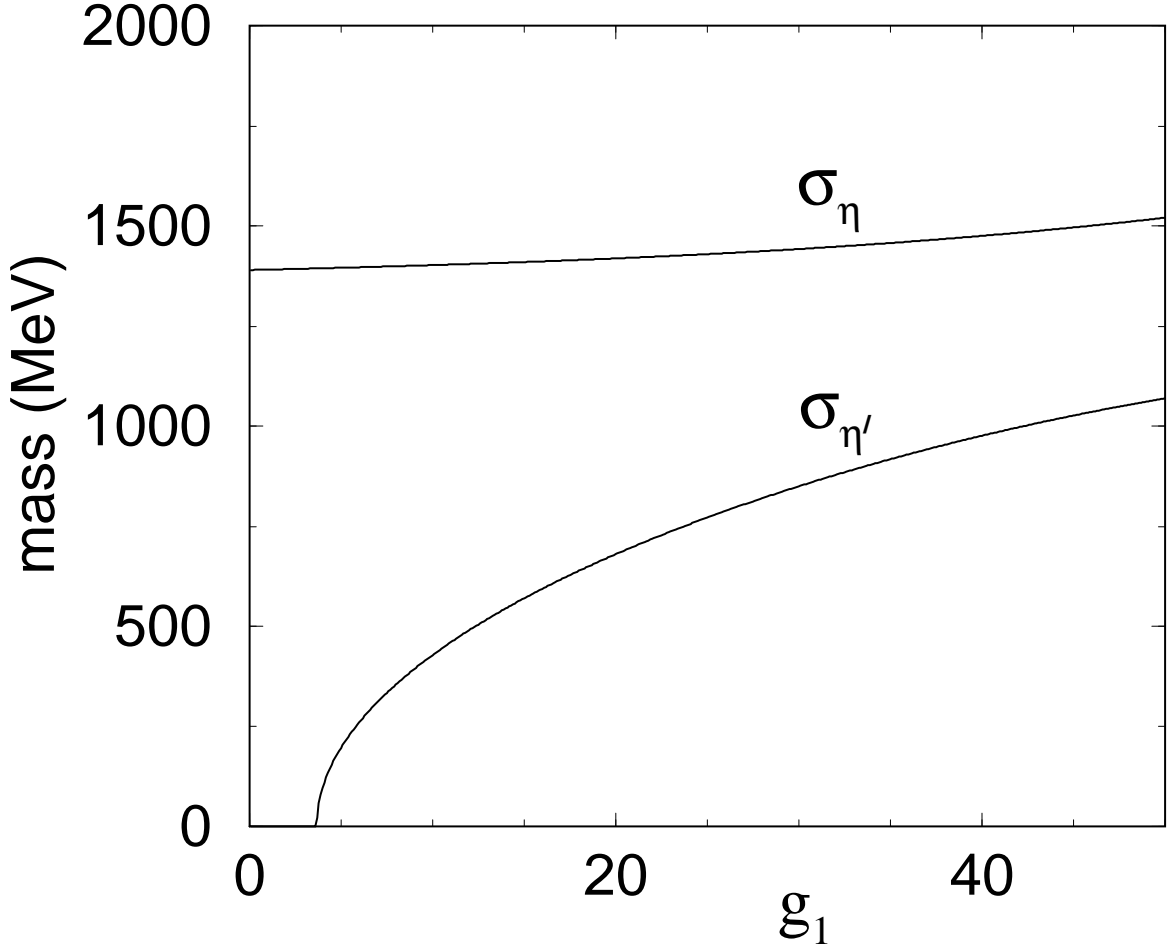}}
\caption{
Masses of the $\sigma_\eta$ and the $\sigma_{\eta '}$, in $MeV$, versus
the coupling $g_1$ of the linear sigma model for three flavors.  Notice
that $m_{\sigma_{\eta '}} = 0$ for $g_1 \sim 4.$; this will be used in
sec. (3.2.3).
}
\end{figure}

Like the $\eta$ and the $\eta'$ mesons,
the $\sigma_\eta$ and the $\sigma_{\eta '}$ mix; the
mixing angles are usually of order $30^0$.  This is close to
ideal mixing, where the $\sigma_{\eta '}$ is primarily $\bar{u}u + \bar{d}d$,
while the $\sigma_\eta$ is primarily $\bar{s} s$.
This mixing helps to understand the differences in the mass spectrum
between the scalars and pseudoscalars.  For the pseudoscalars
the $\eta$ is primarily
an $SU(3)$ octet, and the $\eta'$ primarily an $SU(3)$
singlet; the singlet $\eta'$ is then much heavier than the octet $\eta$
because of the anomaly.  Indeed, the $\eta$ and $\eta'$ are primarily
$SU(3)$ because of the anomaly.
For the scalar particles, instead of being $SU(3)$ eigenstates,
the $\sigma_\eta$ and the $\sigma_{\eta '}$ are much better described
as flavor eigenstates.  Since the $\sigma_{\eta '}$ is (largely)
made up of up and down quarks, then, and the $\sigma_\eta$ is (largely)
made up of strange quarks, it is natural that the for the scalars,
the $\sigma_\eta$ is {\it heavier} than the $\sigma_{\eta '}$.

(I note that in strong
coupling, $g_1 \rightarrow \infty$, $m_{\sigma_{\eta'}}$ approaches
a constant value,
while $m_{\sigma_{\eta}} \sim \sqrt{g_1} \rightarrow \infty$.
This is clearly a special property of taking the up and down quark masses
different from the strange quark mass.)

I identify the $\sigma_{\eta'}$ and the $\sigma_{\eta}$ with the
observed states$^{\ref{pdt}}$ with the same quantum numbers: the $f_0(975)$
and the $f_0(1476)$, respectively.  With the parameters of (\ref{eb11}),
from fig. (12) requiring $m_{\sigma_{\eta'}} = 975 MeV$ fixes
\begin{equation}
m_{\sigma_\eta} \; = \; 1476 MeV \;\; , \;\;
g_1 \; = \; 40. \;\; , \;\;
\mu^2 \; = \; - \, (492 \, MeV)^2 \; .
\label{eb14}
\end{equation}
For these values, the
mixing angle between the $\sigma_{\eta'}$ and the $\sigma_{\eta}$
is $+ 28^o$.  The mass for the $\sigma_\eta$ is a bit high,
$1476 MeV$ instead of $1400 MeV$ for the $f_0$.

Again, the identification of the $\sigma_{\eta'}$ with the
$f_0(975)$ is not evident
(see pg. VII.192 of ref. [\ref{pdt}]):
the $f_0(975)$ has a width of only $\sim 47 \, MeV$, while
the $\sigma_{\eta'}$ should decay into two pions with hearty
abandon.  (The width of the $f_0(1400)$ is large, $150$ to
$400$ MeV.)   Either the narrow width is the result of a complicated
multi-pole structure in the $K\overline{K}$ channel,$^{\ref{penn}}$
or the $f_0(975)$
is not the $\sigma_{\eta'}$, but a $K\overline{K}$
molecule.$^{\ref{kkbar}}$  Indeed, since in this model
the $\sigma_{\eta '}$ is primarily composed of up and down quarks,
and very little from strange quarks, it is impossible to understand
why it should decay into $K\overline{K}$ in the first place.
As shall become clear from our later arguments, for our
purposes all we really need to know that the $\sigma_{\eta'}$ is
not light; {\it i.e.}, $m_{\sigma_{\eta'}} \geq 1 \, GeV$.

So where does the $\sigma$ meson at $600 MeV$,
as we assumed in sec. II, come from?
This is derived from models of
nucleon-nucleon forces, assuming that the force is provided by single
particle exchange.  An isosinglet $0^+$ particle, the $\sigma$, is needed
at $600 MeV$ to provide sufficient attraction.  It is reasonable to
view this $\sigma$ simply as an approximation to two pion exchange.
Notably, there is no such resonance seen
in the phase shifts of $\pi - \pi$ scattering below $1\, GeV$, pg.
VII.37 of ref. [\ref{pdt}].
For the sake of argument, if we do assume that
$m_{\sigma_{\eta'}} = 600 MeV$, we find that the coupling $g_1$ decreases,
to $g_1 = 16.2$, whilst $m_{\sigma_{\eta}} = 1412 MeV$ and
$\mu^2 = + (384 MeV)^2$.

Other fits to the $\sigma_{\eta'}$ have also been given.
Bertsch et al$^{\ref{bert}}$ find that data on kaon decays and
$\pi N \rightarrow \pi \pi N$ scattering can be well fit
with $m_{\sigma_{\eta'}} \sim 800 MeV$.

I now show that understanding where the $\sigma_{\eta '}$ lies is
not merely to interest to afficianadoes of the linear sigma model:
it is crucial to understanding the nature of the phase diagram, as
a function of the current quark mass ${\cal M}$.

\subsubsection{The chiral critical point}

I now turn to the nature of the chiral critical point; the above
classification of the zero temperature spectrum is not required in
this subsection, but will be used in the next.  I deal
exclusively with mean field theory, where the effects of nonzero
temperature are incorporated simply by varying the
mass parameter $\mu^2$.  This is correct in the limit of very high
temperature, but should be qualitatively correct at all temperatures.

I begin with the $SU(3)$ symmetric case when $h_8 = 0$.  For a constant
field $\Sigma_0$, the lagrangian reduces to the potential for $\Sigma_0$,
\begin{equation}
{\cal L} \; = \; - h_0 \, \Sigma_0
\; + \; \frac{1}{2} \, \mu^2 \, \Sigma_0^2
\; - \; \frac{c}{3} \, \Sigma^3_0
\; + \; \frac{g_1}{4} \, \Sigma^4_0 \; .
\label{eb15}
\end{equation}
This model has precisely the same phase diagram as that for the transition
between a liquid and a gas.  For zero background field, $h_0 = 0$, the
instanton coupling ``$c$''
is a cubic invariant, and so drives the transition
first order.  As $h_0$ increases the transition becomes
more weakly first order, until at $h_0 = h_0^{crit}$
the line of first order transitions ends
in a critical point.  For $h_0 > h_0^{crit}$ there is no
true phase transition, just a smooth crossover.

There is a critical point where the
potential in $\Sigma_0 - \Sigma_0^{crit}$
is purely quartic,
\begin{equation}
{\cal L}|_{crit} \; = \; g_1 \left(
\frac{(\Sigma_0 - \Sigma_0^{crit})^4}{4} \right) \; .
\label{eb16a}
\end{equation}
This is determined algebraically, by requiring that about
the critical point, in $\Sigma_0 - \Sigma_0^{crit}$, the first,
second, and third derivatives of the potential vanish.  This
is three conditions, which fixes the values of the magnetic field,
the vacuum expectation value of the $\Sigma_0$ field, and the
mass parameter,
\begin{equation}
h_0^{crit} \; = \; \frac{c^3}{27 \, g^2_1} \;\; , \;\;
\Sigma_0^{crit} \; = \; \frac{c}{2 \, g_1} \;\; , \;\;
\mu^2_{crit} \; = \; \frac{c^2}{3 \, g_1} \; .
\label{eb16}
\end{equation}
The mass spectrum at the critical point is easy to work out by going
back to the original action and recomputing:
$$
m_\pi^2 \; = \; m_K \; =
\; m_\eta^2 \; = \; \frac{c^2}{(9 g_1)} \;\;\; , \;\;\;
m^2_{\eta '} = 10 m^2_\pi \; ,
$$
\begin{equation}
m_{\sigma_{\pi}}^2 \; = \; m_{\sigma_{K}}^2
\; = \; m_{\sigma_{\eta}}^2 \; = \;
\left(7 + 18 \frac{g_2}{g_1} \right) \, m^2_\pi \;\;\; , \;\;\;
m^2_{\sigma_{\eta '}} \; = \; 0 \; .
\end{equation}
Of course the $\sigma_{\eta '}$ is massless because we have deliberately
tuned ourselves to sit a the chiral critical point.  As we work for now
in the limit of $SU(3)$ symmetry, there are two octets of pseudoscalars
and scalar fields.  It certainly is extraordinary to find even an isolated
point where the $\sigma_{\eta '}$ is {\it lighter} than any other field,
even the pions!

Because only one field, the
$\sigma_{\eta'}$, is massless at the chiral critical point,
the similarity to the liquid
gas phase transition extends to the universality class: it is that
of a $Z(2)$ spin system, or the Ising model in three space dimensions.
{\it Thus the chiral critical point is precisely analogous to
critical opalescence in a liquid gas system.}

The classification of the universality class still applies when the
background magnetic field is not $SU(3)$ symmetric,
$h_8 \neq 0$.  In this case I couldn't solve the equations analytically,
and had to resort to numerical methods (which are, however, elementary).
The difficulty is simply that three equations must be set to zero to
find the chiral critical point, for the first, second, and third derivatives
of the potential, in the direction of the massless mode.  (And of course
for the first derivative of the potential, in the other direction.)
By solving (\ref{eb7}) and (\ref{eb8}), I
find that at least for parameters about those
of the previous section, there remains a single, massless field
at the chiral critical point,
the $\sigma_{\eta '}$,
with the universality class that of the Ising model.  Of course for
$h_8 \neq 0$ the $\sigma_{\eta'}$ field does not
remain a pure $SU(3)$ singlet, but mixes to become part octet.

I expect this conclusion to hold for arbitrary values of the quark
masses (background field).  Indeed, I conjecture that in general,
for any critical end point --- be it chiral, deconfining, or whatever ---
that the universality class is always that of the Ising model, which
is $Z(2)$.  I have no profound reason for this, but simply on
algebraic grounds, at best one can get the appropriate derivatives of
the potential to vanish along one direction, but not in more than one
direction; if one tried in more than one direction, there would be more
conditions on the potential than variables to satisfy them.

Indeed, although I did not comment on it at the time,
the identification of the $\sigma_{\eta '}$ as the massless mode of the
chiral critical point could have been anticipated from the results of the
previous section.  Looking at fig. (12), we see that for very small values
of $g_1$, $g_1 \sim 4.$, that the mass of the $\sigma_{\eta '}$ goes to
zero.  At the time, it seemed merely a peculiarity of the fit, of no
greater significance.  And yet it shows that even at zero temperature,
it is possible to obtain a massless $\sigma_{\eta '}$.

Assuming, then, that only the $\sigma_{\eta '}$ is massless at the
chiral critical point, I
can flesh out the entire phase diagram as a function
of ${\cal M}_{up} = {\cal M}_{down}$
versus ${\cal M}_{strange}$, as proposed in fig. (1)
of ref. [\ref{rdp2}].
The basic idea is that along the entire line of
chiral critical points, only the $\sigma_{\eta'}$ is massless, but that
the singlet-octet ratio in the $\sigma_{\eta'}$ changes.
I just showed that at the
$SU(3)$ symmetric point,
${\cal M}_{up}={\cal M}_{down}={\cal M}_{strange}$, the $\sigma_{\eta'}$
is pure singlet.  Decreasing ${\cal M}_{strange}$ to the critical point where
${\cal M}_{strange} =0$ and
${\cal M}_{up}={\cal M}_{down} \neq 0$,
rather obviously the $\sigma_{\eta'}$ must become
entirely strange, $\sigma_{\eta'} \sim \overline{s} s$.

The opposite limit, ${\cal M}_{up}={\cal M}_{down}=0$, is more familiar.
Assume that the chiral phase transition with
two massless flavors is of second order;$^{\ref{col}}$ the universality class
is that of an $O(4)$ critical point.
The chiral phase transition will remain of second order as
${\cal M}_{strange}$ decreases from ${\cal M}_{strange} = \infty$.
This line of second order transitions must end in a special value of
${\cal M}_{strange} = {\cal M}_{strange}^{crit}$;
for ${\cal M}_{strange} < {\cal M}_{strange}^{crit}$,
the chiral transition is like that of three light flavors, and so
first order.  Wilczek$^{\ref{wil}}$ observed that
at ${\cal M}_{strange} =
{\cal M}_{strange}^{crit}$, the chiral transition is in
the universality class of an $O(4)$ {\it tri}-critical point.
I would say that when ${\cal M}_{up}={\cal M}_{down} = 0$ and
${\cal M}_{strange} =
{\cal M}_{strange}^{crit}$, the $\sigma_{\eta '}$ is a pure
$SU(2)$ state, with $\sigma_{\eta '} \sim  \overline{u} u + \overline{d} d$;
further, because
${\cal M}_{up} = {\cal M}_{down} = 0$, the pions are also massless,
so the universality class of this point is not that of a $Z(2)$
critical point, but an $O(4)$ tri-critical point.  This contradicts
my previous argument (that only the $\sigma_{\eta'}$ is massless at
a chiral critical point) because when
${\cal M}_{up} = {\cal M}_{down} = 0$, that is itself the endpoint of
a line of chiral critical points, and thereby special.

The analysis can be further extended to the case where there is a gap
between the
first order transitions for the deconfining and chiral transitions,
but where the chiral transition is of first order for two, massless flavors.
Then fig. (1) of ref. [\ref{col}]
would be modified, with a band of first
order transitions about the axis ${\cal M}_{up} = {\cal M}_{down} = 0$.
These first order transitions would end in chiral critical points,
in the universality class of $Z(2)$ from the presence of
massless $\sigma_{\eta'}$ fields.  However, the latest lattice data appears
to indicate that this is not the case: that for two flavors, the chiral
transition is of second order.

\subsubsection{The chiral critical point and QCD}

To model the effects of
nonzero temperature,
in mean field theory it is only necessary to increase the
mass parameter $\mu^2$ of (\ref{eb2}).  Doing so, the resulting spectrum
--- computed from (\ref{eb6})--(\ref{eb8}) --- is rather unremarkable,
and look very similar to the graph in fig. (1), except that now there
are many more fields about; this is illustrated in fig. (5).
Remember fig. (1) refers to two flavors, including vector mesons,
while fig. (5) is for three flavors, without vector mesons;
also, in fig. (1) I took $m_\sigma(0) = m_{\sigma_{\eta}} =
600 \, MeV$, while in fig. (5) $m_{\sigma_{\eta '}} = 1 \, GeV$.
Yet the qualitative behavior of the states is the same:
the only two states which change significantly with increasing
$\mu^2$ (which corresponds to increasing temperature $T$)
are the pions and the $\sigma_{\eta '}$.
For three flavors, in fig. (5), the
$\sigma_{\eta '}$ comes down in mass from $\sim 1 \, GeV$, while the
pion moves up in mass, until they meet at about $\sim 400 MeV$; then
both increase together.  Since the $\sigma_{\eta '}$ is coming
down in mass from $1 \, GeV$, it is natural to find that it meets up
with the $\pi$ at a much higher mass, $\sim 400 \, MeV$, than before;
if the $\sigma$ starts out at $600 \, MeV$, then it matches up with
the $\pi$ at $\sim 200 \, MeV$.

The limitations of mean field theory are clear.
Including the effects
of nonzero temperature by only changing $\mu^2$ corresponds to
the addition of
an $SU(3)$ symmetric mass term
$\sim \, T^2 \, tr(\Phi^\dagger \Phi)$.
While this type of term is dominant at high temperature, at low temperature
things are surely much more complicated.  Calculating just the thermal
effects to one loop order for the model of (\ref{eb2}) would be
an arduous task.  Nevertheless, we observe that mean field theory
is in qualitative agreement with
more sophisticated analyses in effective
sigma models, instanton models, Nambu Jona-Lasino models, {\it etc.}.
With one exception, these models {\it all} come
to the same conclusion as mean field theory:
in the chiral limit, the chiral transition is
of second order for two flavors,$^{\ref{rdp1},\ref{twofl}}$
first order for three,$^{\ref{threefl}}$ while
for the nonzero quark masses of $QCD$ there is only a smooth
crossover.$^{\ref{rdp1},\ref{twofl},\ref{threefl}}$
I stress that this agreement is rather remarkable, since very
different methods of calculation are being employed.

The one exception is the work of
ref. [\ref{sd}], who find a second order chiral phase transition for three
massless flavors.  These works, however, use an
approximate solution to the Schwinger-Dyson
equations.  This approach does not include the axial anomaly, and so
in this approximation should give a second order phase transition.
Thus the one discrepancy is understandable on general grounds.

We can now use the fit to the zero temperature spectrum to
ask a more detailed question: how far is $QCD$ from the chiral critical
point?  The data of [\ref{col}]
indicates that as a function of
${\cal M} = {\cal M}_{up} = {\cal M}_{down}
= r  \, {\cal M}_{strange}$, $QCD$ is only about a factor of
two from the chiral critical point.

Let us {\it assume} that the chiral phase transition is of first order
for three, massless flavors because of the presence of the instanton
coupling $\sim det(\Phi)$.  Then we assert that
it is {\it very} difficult to understand why
$QCD$ is (relatively) so close to the
chiral critical point.  Varying the current
quark masses is equivalent to varying the background fields $h_0$ and
$h_8$.  We then take (\ref{eb6})--(\ref{eb8}), and
search numerically
for the values of $h_0 = h_0^{crit}$ and $h_8 = h_8^{crit}$ which
give a chiral critical point.  In doing so we require that the ratio
of up to strange quark masses is fixed; from (\ref{eb3}), this implies
that the following ratio of $h_0$ and $h_8$ is fixed to have the
same value as at zero temperature:
\begin{equation}
\frac{(h_0 + 2 h_8)}{(h_0 - h_8)} \; = \; 32 \; .
\label{eb16b}
\end{equation}
By varying $h_0$, $h_8$, $\mu^2$, $\Sigma_0$, and
$\Sigma_8$, and otherwise using the values in (\ref{eb11})
and (\ref{eb14}), I determine where the critical point is.  There
are four equations which must vanish:
two equations of motion, one equation to ensure
the $\sigma_{\eta '}$ is massless, and lastly, one equation
so that the cubic variation of the potential, in the direction of
the $\sigma_{\eta '}$, vanishes.  It is this last equation that makes
the task of determining the location of the critical point so difficult
analytically.  Numerically things are easy, using these four
conditions and (\ref{eb16b}) to determine
$$
h_0^{crit} \; = \; (62 \, MeV)^3 \;\;\; , \;\;\;
h_8^{crit} = (60.4 \, MeV)^3 \;\;\; , \;\;\;
\mu^2_{crit} = (183 \, MeV)^2 \;\;\; , \;\;\;
$$
\begin{equation}
\Sigma_0 = 14.5 \, MeV \;\;\; , \;\;\;
\Sigma_8 = 2.7 \, MeV \; .
\label{eb16c}
\end{equation}
I then compute the ratio between the current quark masses at the
chiral critical point to those in $QCD$:
\begin{equation}
\frac{{\cal M}^{crit}_{up}}{{\cal M}_{up}} \; = \;
\frac{h_0^{crit} - h_8^{crit}}{h_0 - h_8} \; = \;
.01 \; .
\label{eb17}
\end{equation}
Now we easily confess that this ratio, computed in
mean field theory, is crude at best.  Even so, numerical
simulations$^{\ref{col}}$ find that the ratio in (\ref{eb17})
is not $1\%$, but $50\%$ ---
mean field theory is off by almost two orders of magnitude!

For the purposes of discussion, the ratio in (\ref{eb17}) is not significantly
larger for a lighter $\sigma$ meson.  If we take $m_{\sigma_{\eta'}} =
600 MeV$, one finds that
${\cal M}^{crit}_{up}/{\cal M}_{up} = .06$ instead of $.01$.
This is still a long way from the value of $\sim .5$ found by the
Columbia group.$^{\ref{col}}$

Also, I note that
Meyer-Ortmanns, Pirner, and Patkos$^{\ref{mo1}}$ had previously
argued that (in our language) $QCD$ is far from the chiral critical point.
They did not make a precise estimate of this distance, though, as in
(\ref{eb17}).

This leads us to discard our initial assumption.  In the chiral limit of
three, massless flavors, there are two mechanisms for generating a
first order transition.  The first is the presence of the (cubic)
instanton coupling, $\sim det(\Phi)$.$^{\ref{pw},\ref{gold}}$
{}From (\ref{eb17}), this
does not give a phase diagram that is even qualitatively correct.
Therefore, perhaps the second mechanism is at work:$^{\ref{pw}}$
in the chiral limit, the chiral phase transition
is of first order because it is a type of Coleman-Weinberg
transition.$^{\ref{cw},\ref{fi}}$

In mean field theory quartic couplings are fixed and do not change with
temperature.  I use the term Coleman-Weinberg to describe a theory
in which the quartic couplings change significantly with temperature
(in condensed matter physics this is known as ``fluctuation induced'').
This phenomenon can be demonstrated rigorously in
four$^{\ref{cw},\ref{cgs}}$
and $4-\epsilon$$^{\ref{eps}}$ dimensions;
extrapolation to three dimensions, $\epsilon = 1$, may or
may not be valid.  Numerical simulations of
$SU(3) \times SU(3)$ spin models in three dimensions suggest that
they are Coleman-Weinberg like;$^{\ref{kss},\ref{num},\ref{shen}}$
one then has to reconcile this with the expansion up
from two dimensions, which predicts a second order
phase transition.$^{\ref{kss},\ref{rus}}$

(Incidentally,
the claim of Chivukula, Golden, and Simmons$^{\ref{cgs}}$
that the $\beta$-functions in ref. (\ref{eps})
is incorrect is itself
incorrect: the discrepancy is only apparent, from using a different
normalization of the kinetic term.
I thank Y. Shen for pointing this out.)

We propose that the only way for the phase diagram presently observed
in numerical simulations of $2+1$ flavors to be true is if the quartic
couplings of the effective linear sigma model run by large amounts.
In $4-\epsilon$ dimensions,$^{\ref{eps},\ref{shen}}$
in the space of $g_1$ and $g_2$ the couplings tend to run most for
large $g_2$, from large to small values of $g_1$ for approximately constant
$g_2$.  This type of flow is {\it precisely} what is necessary to
obtain a massless $\sigma_{\eta '}$ at the chiral critical point.
As I noted previously, this can be seen even from the parameters of the
model at zero temperature, where a massless $\sigma_{\eta '}$
is obtained by letting $g_1 \rightarrow 4.$.  This hardly proves that
such a speculative scenario is right; but at least couplings flow in
the right direction, with features that are qualitatively correct.

Of course, I confess that using primitive mean field theory is a gross
caricature of reality.  Nevertheless, the question of how far $QCD$ is
from the chiral critical point does not seem to have
been asked previously.  It is surprising that the question of exactly
where and what the $\sigma_{\eta'}$ meson is has such a dramatic effect
on the phase diagram of $QCD$.  If nothing else, while are predictions
are probably baseless, they are precisely testable in numerical simulations
of lattice gauge theory.  Hopefully we won't have to wait until the next
millenium for simulations with dynamical fermions that are good enough
to be believeable.

\subsubsection{Three flavors and DCC's}

Given the crudeness of present day numerical simulations for
$2+1$ flavors, I conclude with wild
speculation: perhaps
$QCD$ is {\it very} close to the chiral critical point.
{\it A priori}, there is no reason why $QCD$ should be very
close to the chiral critical point; there would have to be some
magical conspiracy of mass scales.

But if it is, then even if there is no true phase transition in
$QCD$, there is a natural mechanism for the generation of a large
distance scale at nonzero temperature, from a light $\sigma_{\eta '}$.
As we alluded to in a previous work,$^{\ref{ggp1}}$
this large distance scale {\it might} be used to
produce large domains of Disoriented Chiral Condenstates.
This is not obvious, for even at the chiral critical point,
only the $\sigma_{\eta '}$ becomes massless, not the pions.
The question is then: are large domains of DCC's generated when
massive pions are coupled to a massless (or light)
$\sigma_{\eta'}$?

One way is by the type of isospin zero states constructed
by Horn and Silver$^{\ref{anselm}}$ and discussed recently by
Kowalski and Taylor.$^{\ref{kt}}$  These authors construct a state with
total isospin zero for which distribution in isospin is broad,
like that of a DCC.  The question is then, is it reasonable for such
states to be produced near the chiral critical point?$^{\ref{ggp3}}$


\end{document}